\documentclass[english,prd,superscriptaddress,nofootinbib,preprintnumbers,eqsecnum]{revtex4}
\usepackage[latin1]{inputenc}
\usepackage{graphicx}
\usepackage{amsmath,amsthm,amssymb}
\usepackage{bm}

\usepackage{amsfonts}
\usepackage{dcolumn}

\def\be{\begin{equation}}
\def\ee{\end{equation}}
\def\ba{\begin{eqnarray}}
\def\ea{\end{eqnarray}}
\def\bs{\begin{subequations}}
\def\es{\end{subequations}}

\usepackage{color}

\newcommand{\beqa}{\begin{eqnarray}}
\newcommand{\eeqa}{\end{eqnarray}}

\usepackage{babel}
\begin{document}
\preprint{YITP-14-22}

\title{Cosmological Scaling Solutions for Multiple Scalar Fields}

\author{Takeshi Chiba}
\affiliation{Department of Physics, College of Humanities and Sciences, 
Nihon University, Tokyo 156-8550, Japan}

\author{Antonio De Felice}
\affiliation{Yukawa Institute for Theoretical Physics, Kyoto University, 
Kyoto 606-8502, Japan}

\author{Shinji Tsujikawa}
\affiliation{Department of Physics, Faculty of Science, 
Tokyo University of Science, 
1-3, Kagurazaka, Shinjuku-ku, Tokyo 162-8601, Japan}

\begin{abstract}

The general k-essence Lagrangian for the existence of cosmological 
scaling solutions is derived in the presence of multiple scalar fields 
coupled to a barotropic perfect fluid.
In addition to the scaling fixed point associated with the dynamics 
during the radiation and matter eras, we also obtain a scalar-field 
dominated solution relevant to dark energy and discuss the 
stability of them in the two-field set-up.
We apply our general results to a model of two canonical fields with 
coupled exponential potentials arising in string theory. 
Depending on model parameters and initial conditions, 
we show that the scaling matter-dominated epochs 
followed by an attractor with cosmic acceleration can be 
realized with/without the couplings to scalar fields.
The different types of scaling solutions can be distinguished 
from each other by the evolution of the dark energy equation 
of state from high-redshifts to today.

\end{abstract}

\date{\today}


\maketitle

\section{Introduction}

The scalar fields may have played important 
roles for the expansion history of the Universe. 
The slow-rolling scalar field along a nearly flat potential drives 
the accelerated expansion in the early 
Universe--dubbed inflation \cite{inf}.
The theoretical prediction of inflation for the generation of density 
perturbations from the quantum fluctuation of a scalar degree of 
freedom is consistent with the Cosmic Microwave Background 
(CMB) temperature anisotropies observed by the Planck satellite \cite{Planck}.
The scalar fields can be also responsible for dark energy at the 
expense of having a very light mass of the order of the today's 
Hubble constant $H_0 \simeq 10^{-33}$~eV \cite{quinold,quinpapers}.

The energy scales of scalar fields appearing in particle physics 
are usually much higher than the present cosmological 
density \cite{review}.
The dominance of the field energy density $\rho_x$ 
over the background energy density $\rho_m$ 
in the early Universe after inflation contradicts 
with the successful cosmological sequence 
of the radiation, matter, and accelerated epochs.
If there is a scaling solution where $\rho_x$ is proportional to 
$\rho_m$, however, the Universe can enter the regime in
which the field energy density is sub-dominant to the total 
energy density.

It is well known that a canonical field with the exponential potential 
$V(\phi)=V_0 e^{-\lambda \phi}$ gives rise to scaling 
solutions with $\rho_x/\rho_m={\rm constant}$ \cite{Ratra,clw,Joyce} 
(where the reduced Planck mass $M_{\rm pl}$ is set to unity).
Provided that the constant $\lambda$ satisfies the condition 
$\lambda^2>3(1+w_m)$, where $w_m$ is the equation of state 
of the background barotropic fluid, the solutions approach the scaling 
attractor characterized by the field density parameter 
$\Omega_x=3(1+w_m)/\lambda^2$ \cite{clw}.
In the radiation-dominated epoch there is the bound 
$\Omega_x<0.045$ from the 
Big-Bang-Nucleosynthesis (BBN) \cite{Bean}, so the primordial 
scaling field is compatible with the data for $\lambda>9.4$.

The scaling solutions also arise for non-canonical scalar-field models 
with the Lagrangian $P(\phi,X)$ \cite{kinf,kes}, where $X$ is a kinetic term 
of the field $\phi$. For the existence of scaling solutions the Lagrangian 
is constrained to be in the form $P=Xg(Y)$, where 
$g$ is an arbitrary function in terms of $Y \equiv Xe^{\lambda \phi}$ 
and $\lambda$ is a constant \cite{Piazza,Sami} 
(see also Refs.~\cite{GB,Gomes}). 
This accommodates the case in which the coupling $Q$ between the 
field $\phi$ and the barotropic fluid is present. 
In the absence of the coupling, the field density parameter
relevant to scaling solutions during the radiation or matter eras 
is given by $\Omega_x=3(1+w_m)P_{,X}/\lambda^2$, where 
$P_{,X}=\partial P/\partial X$ \cite{Tsujikawa06,Quartin}.

For the Lagrangian $P=Xg(Y)$, there exists a scalar-field dominated 
solution ($\Omega_x=1$) characterized by the effective 
equation of state $w_{\rm eff}=-1+\lambda^2/(3P_{,X})$ \cite{Tsujikawa06}.
With this solution the accelerated expansion can be realized for 
$\lambda^2/P_{,X}<2$. In this case, however, we do not have 
a physically meaningful scaling solution with $\Omega_x<1$ 
during the radiation and matter eras. 
There are several possible ways of realizing a transition from 
the scaling regime to the epoch of cosmic acceleration.
One is to introduce a canonical single field composed 
by the sum of exponential potentials, e.g.,  
$V(\phi)=V_1 e^{-\lambda_1 \phi}+V_2 e^{-\lambda_2 \phi}$ 
satisfying $\lambda_1^2>3(1+w_m)$ and 
$\lambda_2^2<2$ \cite{bcn} (see also Refs.~\cite{Varun}). 
The joint analysis based on the data of supernovae type Ia, 
CMB, and baryonic acoustic oscillations showed that the 
two slopes are constrained to be 
$\lambda_1>11.7$ and $\lambda_2<0.539$ (95\,\% CL) \cite{CDT}. 

Another way out is to introduce multiple canonical scalar fields 
$\phi_i$ ($i=1,2,\cdots, N$) with the sum of exponential potentials, 
i.e.,  $V=\sum_{i=1}^{N} V_ie^{-\lambda_i \phi_i}$ \cite{Liddle,Malik}. 
In this case there exists a so-called assisted inflationary solution 
characterized by the effective equation of state 
$w_{\rm eff}=-1+\lambda^2/3$, where 
$\lambda \equiv (\sum_{i=1}^N 1/\lambda_i^2)^{-1/2}$.
Even if each field is unable to be responsible for cosmic acceleration 
(i.e., $\lambda_i^2>2$), multiple scalar fields can cooperate 
to give dynamics matching a  single-field solution with $\lambda^2<2$. 
In this model, the dynamics of scaling solutions 
followed by the assisted dark energy attractor 
has been studied in Refs.~\cite{Guo,Blais,Kim,Ohashi,Bruck}.

For the system of multiple fields with the more general 
Lagrangian $P=\sum_{i=1}^{N}X_i\,g(Y_i)$, where $g$ is 
an arbitrary function with respect to 
$Y_i \equiv X_i e^{\lambda_i \phi_i}$, it was shown 
in Ref.~\cite{Ohashi} that the scaling radiation/matter eras 
can be followed by the assisted inflationary attractor. 
However, this analysis does not cover the models 
in which the multiple scalar fields are coupled 
each other \cite{Guo2,tw,ddst}.

Recently, Dodelson {\it et al.} \cite{ddst} presented 
a string-theoretic model described by canonical fields with 
the coupled potential $V=\sum_{i=1}^{2} V_i e^{\alpha_i \phi_1+\beta_i \phi_2}$. 
This multi-field theory follows from a torus compactification 
of an overall volume in the presence of a dilaton field.
Dodelson {\it et al.} showed the existence of not only scaling solutions 
but also solutions with the accelerated expansion relevant to dark energy.
In general there are many scalar fields present in string theory 
(dilaton, moduli, axion), so the action in the Einstein frame contains 
coupled scalar fields after a suitable compactification. 
It is also known that $\alpha'$ corrections to the tree-level string action 
give rise to higher-order derivative terms such as $X^2$ \cite{kinf}.
Then, the general Einstein-frame action with $N$ scalar fields 
can be described by $P(\phi_1,\cdots,\phi_N, X_1,\cdots, X_N)$. 
 
In this paper, we study the construction of the multi-field Lagrangian 
$P(\phi_1,\cdots,\phi_N, X_1,\cdots, X_N)$ 
that possesses scaling solutions. 
For generality we also introduce the couplings 
between the scalar fields and the background barotropic fluid.
The resulting Lagrangian for the existence of scaling solutions 
is surprisingly simple and it covers the multi-field models studied 
in Refs.~\cite{Guo2,tw,ddst} as specific cases.
Moreover, the presence of multiple scalar fields allows the transition 
from the scaling radiation/matter eras to the epoch of cosmic acceleration. 

For the general multi-field scaling Lagrangian we derive the autonomous
equations of motion and the fixed points relevant to dark energy 
as well as the scaling solution.
We show the existence of scalar-field dominated solutions ($\Omega_x=1$)
having the property of assisted inflation. 
For the scaling solution relevant to the dynamics during the radiation and 
matter eras, we also obtain analytic expressions for 
the field density parameter $\Omega_x$ and the field 
equation of state $w_x$. The stability of such fixed points 
will be discussed by considering linear perturbations about them.

As an application of our general scaling Lagrangian, we also study 
the cosmology for a two-field model described by canonical fields 
with the potential $V=\sum_{i=1}^{2} V_i e^{\alpha_i \phi_1+\beta_i \phi_2}$ 
in the absence/presence of couplings between the fields 
and the background barotropic fluid. 
In such a model there exist scaling and accelerated fixed points 
other than those discussed above.
We show how the scaling radiation or matter eras (including the 
$\phi$-matter-dominated-epoch \cite{Amendola}) can be followed by the 
scalar-field dominated attractor with cosmic acceleration.

This paper is organized as follows.
In Sec.~\ref{genesec} we derive the multi-field 
Lagrangian that possesses scaling solutions characterized by 
$\rho_x/\rho_m={\rm constant}$.
In Sec.~\ref{autosec} we obtain autonomous equations and 
some physically important fixed points of the general 
multi-field scaling Lagrangian.
In Sec.~\ref{stasec} the stability of fixed points for 
the scaling solution and the accelerated scalar-field 
dominated point will be discussed. 
In Sec.~\ref{example} we study the cosmological dynamics
for two canonical fields with the potential 
$V=\sum_{i=1}^{2} V_i e^{\alpha_i \phi_1+\beta_i \phi_2}$
in detail.
Sec.~\ref{consec} is devoted to conclusions.

\section{General Multi-field Lagrangian for the Existence of Scaling Solutions}
\label{genesec}

We start with the k-essence model \cite{kinf,kes} described by $N$ 
scalar fields $\phi_i$ ($i=1,2,\cdots,N$) with kinetic energies 
$X_i \equiv -(1/2)g^{\mu \nu}\partial_{\mu}\phi_i \partial_{\nu} \phi_i$, 
where $g^{\mu \nu}$ is the metric tensor.
We also take into account the matter Lagrangian 
${\cal L}_m$ coupled to the scalar fields.
The action in such a system is given by 
\be
S=\int d^4x \sqrt{-g} \left[ \frac12  R
+P(\phi_1,\cdots \phi_N,X_1,\cdots,X_N) \right]
+\int d^4x\,{\cal L}_m (\phi_1,\cdots \phi_N, g_{\mu \nu})\,,
\label{actionkes}
\ee
where $g$ is the determinant of the metric $g_{\mu \nu}$, 
$P$ is the Lagrangian of multiple scalar fields, and 
${\cal L}_m$ is the matter Lagrangian. 
We use the unit $M_{\rm pl}=1$.

We introduce couplings between the scalar fields 
and matter in the form
\be
Q_i (\phi_i)=-\frac{1}{\rho_m} \frac{1}{\sqrt{-g}}
\frac{\partial {\cal L}_m}{\partial \phi_i} \qquad
(i=1,2,\cdots,N),
\label{couplings}
\ee
where $\rho_m$ is the energy density of matter. 
The couplings $Q_i$ are analogous to those appearing in 
the coupled quintessence scenario discussed 
in Refs.~\cite{Amendola,Piazza}. 
In Brans-Dicke (BD) theory \cite{Brans}, for example, the BD scalar 
field couples to non-relativistic matter with constant 
couplings $Q_i$ in the Einstein frame \cite{dn,Khoury,Yoko,cy}. 
In such theory the radiation does not couple to the BD scalar 
because it is traceless. 
In the following we consider a barotropic perfect fluid 
with the constant equation of state $w_m$ coupled to the fields $\phi_i$.
Although we basically have the matter-dominated epoch 
in mind, Our analysis also covers the radiation era 
by setting $Q_i$ zero for radiation.  
In the rest of the paper we assume that $w_m$ is 
in the range $0 \leq w_m <1$.

\subsection{Background equations of motion}

Let us study the cosmology on the flat Friedmann-Lema\^{i}tre-Robertson-Walker 
background with the line-element $ds^2=-dt^2+a^2(t) \delta_{ij}dx^idx^j$, 
where $a(t)$ is the scale factor with cosmic time $t$.
Varying the action (\ref{actionkes}) with respect to $\phi_i$, 
we obtain the following equations of motion 
for $N$ scalar fields
\be
\frac{d}{dt} ( \dot{\phi}_iP_{,X_i} )+3H \dot{\phi}_iP_{,X_i}
-P_{,\phi_i}=-Q_i\rho_m\,,\qquad
(i=1,2,\cdots,N)\,,
\label{fieldeq1}
\ee
where a dot represents a derivative with respect to $t$, 
$H \equiv \dot{a}/a$, $P_{,X_i} \equiv \partial P/\partial X_i$, 
and $P_{,\phi_i} \equiv \partial P/\partial \phi_i$. 
In the following we assume that the couplings $Q_i$ 
are constant\footnote{Along the line of Ref.~\cite{Quartin},
it is also possible to generalize the analysis 
to the case in which the couplings depend on scalar fields.}.

Multiplying the term $\dot{\phi}_i$ for Eq.~(\ref{fieldeq1}) 
and summing up each equation from $i=1$ to $i=N$,
it follows that 
\be
\dot{\rho}+3H (1+w_x) \rho=-\sum_{i=1}^{N} 
Q_i \rho_m \dot{\phi}_i\,,
\label{rhoeq}
\ee
where 
\be
\rho \equiv \sum_{i=1}^N 2X_i P_{,X_i}-P\,,\qquad
w_x \equiv P/\rho\,.
\ee
Due to the energy conservation the barotropic 
perfect fluid obeys
\be
\dot{\rho}_m+3H (1+w_m) \rho_m
=\sum_{i=1}^{N} Q_i \rho_m \dot{\phi}_i\,.
\label{rhomeq}
\ee
{}From the Einstein equations we obtain
\beqa
& & 3 H^2=\rho+\rho_m\,,\label{Ein1}\\
& & 2 \dot{H}=-(1+w_x)\rho-(1+w_m)\rho_m\,.
\label{dotH}
\eeqa
We also define the effective equation of state 
\be
w_{\rm eff} \equiv -1-\frac{2\dot{H}}{3H^2}\,.
\label{weff}
\ee
{}From Eqs.~(\ref{Ein1}) and (\ref{dotH}) we can express
$w_{\rm eff}$ in the form
\be
w_{\rm eff}=w_m+(w_x-w_m)\Omega_x\,,\qquad {\rm where}
\qquad \Omega_x \equiv \frac{\rho}{\rho+\rho_m}\,.
\label{weff2}
\ee

\subsection{Derivation of the multi-field scaling Lagrangian}

Let us derive the Lagrangian that gives rise to 
scaling solutions characterized by $\rho/\rho_m=C$, 
where $C$ is a non-zero constant.
This demands the condition 
\be
\frac{\rho'}{\rho}=\frac{\rho_m'}{\rho_m}\,,
\ee
where a prime represents a derivative with 
respect to the e-folding number ${\cal N}=\ln a$ (having the relation $d{\cal N}/dt=H$).
On using the relations (\ref{rhoeq}) and (\ref{rhomeq}), it follows that 
\be
\sum_{i=1}^N Q_i \phi_i'=3(w_m-w_x)\Omega_{x}\,,
\label{Qre}
\ee
where
\be
(\ln \rho)^{'}=(\ln \rho_m)^{'}
=-3(1+w_{\rm eff})\,.
\label{rhodef}
\ee
In the scaling regime where $w_x$ and $\Omega_x$ are constant, 
we can integrate Eq.~(\ref{rhodef}) to give 
\be
\rho \propto \rho_m \propto a^{-3(1+w_{\rm eff})}\,.
\ee
Since $H^2 \propto \rho \propto a^{-3(1+w_{\rm eff})}$,
the scale factor and the Hubble parameter evolve, respectively, as
\be
a \propto t^{2/[3(1+w_{\rm eff})]}\,,\qquad
H=\frac{2}{3(1+w_{\rm eff})\,t}\,.
\label{aH}
\ee

Each term on the left-hand side (lhs) of Eq.~(\ref{Qre}) can be written in the form 
$Q_i \phi_i'=3(w_m-w_x)\Omega_{x}r_i$, where 
$\sum_{i=1}^N r_i=1$. 
The ratios $r_i$ should be constant in the scaling regime. 
On using Eq.~(\ref{weff2}), it follows that 
\be
\phi_i'=\frac{3(w_m-w_{\rm eff})r_i}{Q_i}\,.
\label{phi12}
\ee
Then the field kinetic energies have the dependence
$X_i=H^2\phi_i'^2/2 \propto H^2 \propto \rho$, so that 
\be
(\ln X_i)^{'}=-3(1+w_{\rm eff})\,.
\label{Xi}
\ee
Moreover, since $w_x=P/\rho$ is constant in the scaling regime, 
the pressure $P$ has the same dependence as $\rho$. 
Hence we have 
\be
(\ln P)^{'}=\sum_{i=1}^{N}
\left[ \frac{\partial \ln P}{\partial \phi_i} \phi_i'+
\frac{\partial \ln P}{\partial \ln X_i}(\ln X_i)^{'}
\right]=-3(1+w_{\rm eff})\,.
\label{lnp}
\ee
Substitution of Eqs.~(\ref{phi12}) and (\ref{Xi}) 
into Eq.~(\ref{lnp}) reads
\be
\sum_{i=1}^N \left( \frac{\partial \ln P}{\partial \ln X_i}
-\frac{1}{\lambda_i} \frac{\partial \ln P}{\partial \phi_i} \right)
=1\,,
\label{Pdef}
\ee
where 
\be
\lambda_i \equiv \frac{(1+w_{\rm eff})Q_i}
{(w_m-w_{\rm eff})r_i}\,.
\label{lambdai}
\ee
Integrating Eq.~(\ref{Pdef}), we obtain the Lagrangian
\be
P=e^{-\lambda_1 \phi_1} g(Y_1,\cdots ,Y_N,Z_1,\cdots ,Z_{N-1})\,,
\label{scalinglag}
\ee
where $g$ is an arbitrary function with respect to
\be
Y_i \equiv X_i e^{\lambda_1\phi_1}\,,\qquad
Z_{i} \equiv \phi_{i+1}-\frac{\lambda_1}{\lambda_{i+1}}\phi_1\,.
\label{Zi}
\ee
In the presence of the field $\phi_1$ alone we have 
$Y_i=0$ ($i \geq 2$) and $Z_i=0$ ($i \geq 1$) 
(because $r_i \to 0$ and $\lambda_i \to \infty$ for $i \geq 2$), in which case 
the Lagrangian (\ref{scalinglag}) reduces to 
$P=e^{-\lambda_1 \phi_1} g(X_1 e^{\lambda_1\phi_1})$. 
This corresponds to the single-field scaling 
Lagrangian\footnote{Refs.~\cite{Piazza,Sami} derived  
the Lagrangian in the form $P=X_1g(X_1 e^{\lambda_1\phi_1})$, but this is 
equivalent to $P=e^{-\lambda_1 \phi_1} g(X_1 e^{\lambda_1\phi_1})$.}
derived in Refs.~\cite{Piazza,Sami}.

In the scaling regime we have $\phi_i'=3(1+w_{\rm eff})/\lambda_i={\rm constant}$ 
from Eqs.~(\ref{phi12}) and (\ref{lambdai}). 
Integrating this equation and using Eq.~(\ref{aH}), 
each field evolves as  
\be
\phi_i=\phi_{i}^{(0)}+\frac{2}{\lambda_i} 
\ln \left( \frac{t}{t_0} \right)\,,
\label{phiso}
\ee
where $\phi_{i}^{(0)}$ is the field value at $t=t_0$. 
Along this solution the quantities $Y_i$ and $Z_i$ behave as
\beqa
Y_i &=&\frac12 \dot{\phi}_i^2 e^{\lambda_1 \phi_1} 
\propto t^{-2}\,t^2={\rm constant}\,,\\
Z_i &=& \phi_{i+1}^{(0)}-\frac{\lambda_1}{\lambda_{i+1}} \phi_1^{(0)}
={\rm constant}\,.
\eeqa
Hence the function $g$ remains constant for the scaling 
solution discussed above.

\subsection{Specific scaling models}

The Lagrangian (\ref{scalinglag}) is general enough to cover 
a wide variety of multi-field models having scaling solutions.

Let us first consider the two-field model described by 
\be
g(Y_1,Y_2,Z_1)=Y_1+Y_2-V_1 e^{\mu_1Z_1}-V_2 e^{\mu_2Z_1}\,,
\label{Dodelson}
\ee
where $V_i$ and $\mu_i$ ($i=1,2$) are constants.
Then the Lagrangian (\ref{scalinglag}) reads
\be
P=X_1+X_2-V_1 e^{\alpha_1\phi_1+\mu_1\phi_2}
-V_2e^{\alpha_2\phi_1+\mu_2\phi_2}\,,
\label{Plag}
\ee
where 
\be
\alpha_1 \equiv -\lambda_1(1+\mu_1/\lambda_2)\,\qquad
\alpha_2 \equiv -\lambda_1(1+\mu_2/\lambda_2)\,.
\label{alvalues}
\ee
This is the model appearing in string theory \cite{ddst}. 

Under the scaling transformation with a parameter ${\cal A}$:
\beqa
& &
\phi_1 \rightarrow \phi_1 -2\frac{\mu_1-\mu_2}
{\alpha_2 \mu_1-\alpha_1 \mu_2}{\cal A}=
\phi_1+\frac{2{\cal A}}{\lambda_1}\,,\\
& &
\phi_2 \rightarrow \phi_2 +2\frac{\alpha_1-\alpha_2}
{\alpha_2\mu_1-\alpha_1\mu_2}{\cal A}
=\phi_2+\frac{2{\cal A}}{\lambda_2}\,, 
\label{scaling}
\eeqa
the sum of the two potentials
 in Eq.~(\ref{Plag}) transforms homogeneously: 
 $V \rightarrow V e^{-2{\cal A}}$ \cite{ddst}.
The scaling solution corresponds to the case in which 
each term in Eq.~(\ref{Plag}) evolves in the same 
way as $H^2 \propto t^{-2}$.
On using the transformation (\ref{scaling}), we find that 
the scaling solution (\ref{phiso}) has such a property. 
We stress that the variables $\lambda_i$ ($i=1,2$) are
more fundamental than $\alpha_i$ to determine the property 
of scaling solutions. 
The variables $\mu_1$ and $\mu_2$ also matter when we 
discuss the stability of the solution (\ref{phiso}).

We can also consider more general models with two canonical 
fields described by 
\be
g(Y_1,Y_2,Z_1)=Y_1+Y_2-h(Z_1)\,,
\ee
where $h(Z_1)$ is an arbitrary function with respect to $Z_1$. 
In this case the field has a potential of the form 
\be
V(\phi_1,\phi_2)=e^{-\lambda_1 \phi_1}\,
h \left( \phi_2-\frac{\lambda_1}{\lambda_2}\phi_1 \right)\,.
\label{twofieldpo}
\ee
It is instructive to define the field $\sigma$ along the scaling solution 
and the orthogonal field $s$ in terms of their velocities as \cite{gordon,tegmark}
\beqa
\dot{\sigma} &=&
(\cos \theta) \dot{\phi}_1
+(\sin \theta) \dot{\phi}_2\,,\\
\dot{s} &=& 
-(\sin \theta) \dot{\phi}_1
+(\cos \theta) \dot{\phi}_2\,.
\label{rotation}
\eeqa
where $\cos \theta \equiv \dot{\phi}_1/\sqrt{\dot{\phi}_1^2
+\dot{\phi}_2^2}$ and 
$\sin \theta \equiv \dot{\phi}_2/\sqrt{\dot{\phi}_1^2
+\dot{\phi}_2^2}$.
The scaling solution (\ref{phiso}) obeys the following relations
\be
\cos \theta=\frac{\lambda}{\lambda_1}\,,\qquad
\sin \theta=\frac{\lambda}{\lambda_2}\,,
\ee
where $1/\lambda^2=1/\lambda_1^2+1/\lambda_2^2$.
Since $\theta$ is constant, the trajectory in field space
is a straight line. The velocity $\dot{s}$ of the orthogonal   
field vanishes by definition. 
Then the scaling solution in the two-field system 
can be described by an effective single-field 
trajectory with 
\be
\dot{\sigma}=\frac{2}{\lambda t}\,,\qquad
\sigma=\sigma^{(0)}+\frac{2}{\lambda} 
\ln \left( \frac{t}{t_0} \right)\,,
\label{dotsigma}
\ee
where $\sigma^{(0)}$ is a constant. 
The effective slope $\lambda$, which is the combination 
of $\lambda_1$ and $\lambda_2$, is crucial to determine
the property of scaling solutions.
This is also the case for more general non-canonical 
fields described by the Lagrangian (\ref{scalinglag}). 
We shall study this issue in detail in the next section.

The above prescription can be generalized to the 
$N$-scalar field models with 
\be
g(Y_1,\cdots,Y_N,Z_1,\cdots ,Z_{N-1})
=\sum_{i=1}^{N}Y_i-h(Z_1,\cdots ,Z_{N-1})\,,
\ee
where $h$ is an arbitrary function in terms of $Z_1,\cdots,Z_{N-1}$.
We introduce the field $\sigma$ in the form
\be
\dot{\sigma}=\sum_{i=1}^{N} \beta_i \dot{\phi}_i\,,\qquad
\beta_i \equiv \frac{\dot{\phi}_i}{\sqrt{\sum_{i=1}^{N} \dot{\phi}_i^2}}\,.
\ee
The scaling solution (\ref{phiso}) satisfies
\be
\beta_i=\frac{\lambda}{\lambda_i}\,,\qquad {\rm where}
\qquad
\frac{1}{\lambda^2}=\sum_{i=1}^{N} 
\frac{1}{\lambda_i^2}\,.
\label{lade}
\ee
Then, the scaling solution in the $N$-field system obeys the effective 
single-field trajectory given by Eq.~(\ref{dotsigma}).

\section{Autonomous Equations and Fixed Points for the Multi-field 
Scaling Lagrangian}
\label{autosec}

In this section we derive the dynamical equations of motion and 
fixed points for the general Lagrangian (\ref{scalinglag}) 
with $N$ scalar fields.
We also take into account a barotropic perfect fluid coupled to 
the fields $\phi_i$ with $Q_i$ given in Eq.~(\ref{couplings}).

\subsection{Autonomous equations}

In order to study the autonomous equations of motion, 
we introduce the following dimensionless variables:
\be
x_i\equiv\frac{\dot\phi_i}{\sqrt{6}H}\,,\qquad
y\equiv \frac{e^{-\lambda_1\phi_1/2}}{\sqrt{3}H}\,, 
\qquad (i=1,2,\cdots,N)\,,
\label{ydef}
\ee
and $Z_i$ defined in Eq.~(\ref{Zi}) with $i=1,2,\cdots,N-1$.
It is convenient to notice the following relations
\beqa
&& X_i=3H^2x_i^2\,,\qquad Y_i=x_i^2/y^2\,,\qquad
P_{,X_i}=g_{,Y_i}\,,\nonumber \\
&& 
P_{,\phi_1}=3\lambda_1H^2 \left[ 
\sum_{i=1}^{N} x_i^2 g_{,Y_i}
-gy^2-y^2 \sum_{i=1}^{N-1} \frac{g_{,Z_i}}{\lambda_{i+1}} 
\right]\,,\qquad
P_{,\phi_i}=3H^2y^2g_{,Z_{i-1}}\,,\qquad (i \geq 2),\nonumber \\
& & \Omega_x=\sum_{i=1}^{N} 2x_i^2 g_{,Y_i}-gy^2
\label{omegax}\,,\qquad
w_x=\frac{gy^2}{\sum_{i=1}^{N} 2x_i^2 g_{,Y_i}-gy^2}\,.
\eeqa
{}From Eq.~(\ref{omegax}) we find
\beqa
& & gy^2=\Omega_x w_x\,,\label{Omew0}\\
& & \sum_{i=1}^{N} 2x_i^2 g_{,Y_i}= \Omega_x(1+w_x)\,.
\label{Omew}
\eeqa

On using Eqs.~(\ref{rhoeq})-(\ref{dotH}), we obtain the following 
autonomous equations
\beqa
x_1' &=& \frac32 x_1 \left[ w_m+(w_x-w_m)\Omega_x-1 \right] \nonumber \\
&&-\frac{1}{g_{,Y_1}} \left[ x_1 g_{,Y_1}'-\frac{\sqrt{6}}{2}\lambda_1
\left\{ \sum_{i=1}^{N} x_i^2 g_{,Y_i}
-gy^2-y^2 \sum_{i=1}^{N-1} \frac{g_{,Z_i}}{\lambda_{i+1}}
\right\}+\frac{\sqrt{6}}{2}Q_1 (1-\Omega_x) \right]\,,
\label{x1}\\
x_i' &=& \frac32 x_i \left[ w_m+(w_x-w_m)\Omega_x-1 \right] 
-\frac{1}{g_{,Y_i}} \left[ x_i g_{,Y_i}'-\frac{\sqrt{6}}{2}y^2 g_{,Z_{i-1}}
+\frac{\sqrt{6}}{2} Q_i(1-\Omega_x) \right]\,,\qquad (i=2,3,\cdots,N)\,,
\label{x2}\\
y'&=&y \left[ -\frac{\sqrt{6}}{2} \lambda_1 x_1+\frac32
\left\{ w_m+(w_x-w_m)\Omega_x+1 \right\} \right]\,,
\label{y}\\
Z_i'&=&\sqrt{6}\left(x_{i+1}-\frac{\lambda_1}{\lambda_{i+1}}x_1\right)
\,,\qquad (i=1,2,\cdots,N-1)\,.
\label{z1}
\eeqa
This system is described by $2N$ differential equations of motion.
For a given $g(Y_1,\cdots Y_N,Z_1,\cdots,Z_{N-1})$, 
the cosmological evolution is known 
by solving Eqs.~(\ref{x1})-(\ref{z1}).

\subsection{Fixed points of assisted inflation and the scaling solution}

Let us search for fixed points with constant values of 
$x_i$, $y$, and $Z_i$. We caution that there are some 
cases in which $y$ and $Z_i$ vary in time, while keeping 
the variables such as $x_i$ and $gy^2$ remain constant. 
We will see such examples in Sec.~\ref{example}.

Since $g_{,Y_1}$ and $g_{,Y_2}$ are constants, 
it follows that $g_{,Y_1}'=0$ and $g_{,Y_2}'=0$. 
The fixed points relevant to scaling solutions discussed 
in Sec.~\ref{genesec} correspond to 
\be
y \neq 0\,.
\ee

{}From Eqs.~(\ref{y}) and (\ref{z1}) we obtain
\be
x_i=\frac{\sqrt{6}}{2\lambda_i}(1+w_{\rm eff})\,,\quad
{\rm for} \quad i=1,2,\cdots,N\,,
\label{x1x2}
\ee
where $w_{\rm eff}$ is given by Eq.~(\ref{weff2}).
Substituting Eq.~(\ref{x1x2}) into Eqs.~(\ref{x1})-(\ref{x2})
and using the relations (\ref{Omew0})-(\ref{Omew}), 
we obtain the following relations
\beqa
y^2 \sum_{i=1}^{N-1} \frac{g_{,Z_i}}{\lambda_{i+1}}
&=&
\frac12 \Omega_x (1-w_x)
-\frac{3g_{,Y_1}}{2\lambda_1^2}(1-w_{\rm eff}^2)
-\frac{Q_1}{\lambda_1} (1-\Omega_x)\,, \label{yeq1}\\
y^2 g_{,Z_{i-1}}
&=& 
\frac{3g_{,Y_i}}{2\lambda_i} \left( 
1-w_{\rm eff}^2 \right)+
Q_i (1-\Omega_x)
\,,\qquad \quad  (i=2,3,\cdots,N).
\label{yg}
\eeqa
Changing each subscript $i$ in Eq.~(\ref{yg}) to $i+1$, dividing 
the resulting equation by $\lambda_{i+1}$, and summing up each 
equation from $i=1$ to $i=N-1$, it follows that 
\be
y^2 \sum_{i=1}^{N-1} \frac{g_{,Z_i}}{\lambda_{i+1}}
=\sum_{i=1}^{N-1} \left[ \frac{3g_{,Y_{i+1}}}
{2\lambda_{i+1}^2} \left( 1-w_{\rm eff}^2 \right)+
\frac{Q_{i+1}}{\lambda_{i+1}} (1-\Omega_x)\right]\,.
\label{yeq2}
\ee
{}From Eqs.~(\ref{yeq1}) and (\ref{yeq2}) we find
\be
\Omega_x (1-w_x)-(1-w_{\rm eff}^2)\sum_{i=1}^{N} 
\frac{3g_{,Y_i}}{\lambda_i^2} 
=2q (1-\Omega_x)\,,
\label{Omerela}
\ee
where 
\be
q \equiv \sum_{i=1}^{N} \frac{Q_i}{\lambda_i}\,.
\label{qdef}
\ee
Substitution of Eqs.~(\ref{x1x2}) into Eq.~(\ref{Omew}) reads
\be
\sum_{i=1}^{N} 
\frac{3g_{,Y_i}}{\lambda_i^2}=
\frac{\Omega_x (1+w_x)}{(1+w_{\rm eff})^2}\,.
\label{gyi}
\ee
Plugging Eqs.~(\ref{gyi}) and (\ref{weff2}) into 
Eq.~(\ref{Omerela}), we obtain
\be
(\Omega_x-1) \left[ (1+q)(w_m-w_x)\Omega_x
-q(1+w_m) \right]=0\,.
\ee

This shows that there exist the following two fixed points:
\begin{itemize}
\item{Point A: a scalar-field dominated point with}
\beqa
\Omega_x=1\,.
\eeqa
\item{Point B: a scaling solution with}
\beqa
\Omega_x=\frac{q(1+w_m)}
{(w_m-w_x)(1+q)}\,.
\label{Omexscaling}
\eeqa
In the single-field limit $q \to Q_1/\lambda_1$, this recovers the 
result $\Omega_x=Q_1(1+w_m)/[(w_m-w_x)(\lambda_1+Q_1)]$ 
derived in Ref.~\cite{Tsujikawa06}.
\end{itemize}

In the following we discuss the above two fixed points in more detail.

\subsubsection{Point A: the scalar-field dominated point}

Substituting $\Omega_x=1$ into Eqs.~(\ref{weff2}) and (\ref{Omew0}), 
it follows that $w_{\rm eff}=w_{x}=gy^2$.
On using Eqs.~(\ref{x1x2}), (\ref{gyi}), and (\ref{yg}) 
we obtain the relations 
\beqa
& & g = \frac{\sqrt{6}\lambda_1 x_1-3}{3y^2}\,,
\label{sca1} \\
& &\sum_{i=1}^{N} \frac{g_{,Y_i}}{\lambda_i^2}
=\frac{1}{\sqrt{6}\lambda_1 x_1}\,,
\label{sca2}\\
& & \frac{g_{,Z_i}}{g_{,Y_{i+1}}} =
\frac{\lambda_1 x_1 (\sqrt{6}-\lambda_1 x_1)}
{\lambda_{i+1} y^2}\,.
\label{sca3}
\eeqa
{}From Eq (\ref{sca2}) we have 
\be
x_1=\frac{\lambda^2}{\sqrt{6}\lambda_1}\,,
\label{x1so}
\ee
where 
\be
\frac{1}{\lambda^2} \equiv 
\sum_{i=1}^{N} \frac{g_{,Y_i}}{\lambda_i^2}\,.
\label{lamdef}
\ee
Substituting Eq.~(\ref{x1so}) into Eq.~(\ref{sca1}), 
the effective equation of state $w_{\rm eff}=gy^2$ 
can be expressed as
\be
w_{\rm eff}=-1+\frac{\lambda^2}{3}\,.
\label{weffA}
\ee
The cosmic acceleration occurs for $\lambda^2<2$. 
Even if each field does not lead to the cosmic acceleration, 
the presence of multiple fields can do so by reducing the 
value of $\lambda^2$ relative to $\lambda_i^2/g_{,Y_i}$.
This is the phenomenon of assisted 
inflation, which is known to occur for canonical scalar 
fields with exponential potentials \cite{Liddle}. 
The above argument shows that the assisted inflationary 
mechanism is also present for the general multi-field scaling 
Lagrangian (\ref{scalinglag}).
For a given function $g$ we can solve Eqs.~(\ref{sca1})-(\ref{sca3})
for $x_i$, $y$, and $Z_i$ by noting the relations 
$Y_i=x_i^2/y^2$ and $Y_j=Y_i \lambda_i^2/\lambda_j^2$. 

For example, let us consider the two-field model 
\be
g(Y_1,Y_2,Z_1)=Y_1+Y_2-V_1-V_2 e^{-\lambda_2 Z_1}\,,
\label{doubleex}
\ee
which corresponds to $\mu_1=0$ and $\mu_2=-\lambda_2$ 
in Eq.~(\ref{Dodelson}).
This model is characterized by two canonical fields 
with exponential potentials: 
$P=X_1+X_2-V_1e^{-\lambda_1 \phi_1}
-V_2e^{-\lambda_2 \phi_2}$ \cite{Liddle}.
Solving Eqs.~(\ref{sca1})-(\ref{sca3}) in this case, 
we obtain $x_1=\lambda^2/(\sqrt{6}\lambda_1)$ and 
\be
y^2=\frac{1}{V_1} \left( 1-\frac{\lambda^2}{6} \right) 
\frac{\lambda^2}{\lambda_1^2}\,,\qquad
e^{-\lambda_2 Z_1}=\frac{\lambda_1^2}{\lambda_2^2}
\frac{V_1}{V_2}\,,
\ee
where $1/\lambda^2=1/\lambda_1^2+1/\lambda_2^2$. 
The density parameters of the fields $\phi_1$ and $\phi_2$ 
are given by 
$\Omega_{x_1}=x_1^2+V_1 y^2=\lambda^2/\lambda_1^2$ and 
$\Omega_{x_2}=x_2^2+V_2 y^2e^{-\lambda_2 Z_1}
=\lambda^2/\lambda_2^2$, respectively, so that 
$\Omega_{x_1}/\Omega_{x_2}=\lambda_2^2/\lambda_1^2$.
The field with a smaller value of $\lambda_i$ has a lager energy 
fraction than that of another field.

\subsubsection{Point B: the scaling solution}

Substituting Eq.~(\ref{Omexscaling}) into Eq.~(\ref{weff2}), 
the effective equation of state for the scaling solution is given by 
\be
w_{\rm eff}=\frac{w_m-q}{1+q}\,,
\label{weffscaling}
\ee
where $q$ is defined by Eq.~(\ref{qdef}).
For non-relativistic matter ($w_m=0$) the presence of the 
couplings $Q_i$ can lead to cosmic acceleration for 
$q>1/2$. {}From Eqs.~(\ref{x1x2}), (\ref{gyi}) and (\ref{Omew0}), 
we obtain the following relations
\beqa
& & x_1=\frac{\sqrt{6}}{2\lambda_1} 
\frac{1+w_m}{1+q}\,,
\label{scaling1} \\
& & \frac{1}{\lambda^2}
=\frac{\Omega_x(1+w_x)(1+q)^2}{3(1+w_m)^2}\,,
\label{scaling121}\\
& & gy^2=\frac{w_x(1+w_m)q}{(w_m-w_x)(1+q)}\,,
\label{scaling2}
\eeqa
where $\lambda^2$ is defined in Eq.~(\ref{lamdef}).
Using Eqs.~(\ref{Omexscaling}) and (\ref{scaling121}), 
$\Omega_x$ and $w_x$ can be written as
\beqa
\Omega_x &=&
\frac{3(1+w_m)+q(1+q)\lambda^2}{(1+q)^2\lambda^2}\,,
\label{Omex} \\
w_x &=& 
\frac{3w_m(1+w_m)-q(1+q)\lambda^2}
{3(1+w_m)+q(1+q)\lambda^2}\,.
\label{wx}
\eeqa

In the $q \to 0$ limit we have $\Omega_x= 
3(1+w_m)/\lambda^2$ and $w_{\rm eff}=w_x=w_m$, 
so the energy density of the scalar fields scale 
as that of the background fluid. 
For the physically meaningful scaling solution characterized 
by $\Omega_x<1$, we require that $\lambda^2>3(1+w_m)$.
This condition is incompatible with that for the cosmic acceleration 
of the fixed point A ($\lambda^2<2$).
However, if $\lambda_1^2$ is larger than the order of 1 and
$\lambda_2^2$ is smaller than $O(1)$, there is a possibility 
that the field $\phi_1$ is in the scaling regime during the 
radiation and matter eras and that the field $\phi_2$ leads
to the acceleration of the Universe at late times.
We will discuss these cases in Sec.~\ref{example}.

In the presence of the couplings $Q_i$, we also have another 
possibility to give rise to the accelerated expansion by the scaling 
fixed point B with $\Omega_x \simeq 0.7$ for $q>1/2$. 
In this case the standard matter era is also replaced by 
another solution (fixed point C discussed below). 
In the single-field scenario it was shown 
that the scaling accelerated solution B preceded by the 
modified matter era is difficult to be realized 
for a vast class of scaling models \cite{Quartin}.
In Sec.~\ref{example} we mention this possibility for 
the two-field model (\ref{Dodelson}).

Provided that the function $g$ is given, we can solve 
Eqs.~(\ref{scaling2}) and (\ref{yg}) for $y$ and $Z_1$. 
For the model (\ref{doubleex}) with 
$w_m=0$, it follows that 
\beqa
y^2 &=& \frac{2Q_2 \lambda_2 (3+Q_1 \lambda_1)
+[3+2Q_1(Q_1+\lambda_1)]\lambda_2^2-6Q_1 \lambda_1}
{2V_1 [Q_1 \lambda_2+(Q_2+\lambda_2)\lambda_1]^2}\,,\\
e^{\lambda_2 Z_1} &=& 
\frac{2Q_2 \lambda_2 (3+Q_1 \lambda_1)+
[3+2Q_1 (Q_1+\lambda_1)]\lambda_2^2-6Q_1\lambda_1}
{2Q_1 \lambda_1 (3+Q_2 \lambda_2)+
[3+2Q_2 (Q_2+\lambda_2)]\lambda_1^2-6Q_2\lambda_2}
\frac{V_2}{V_1}\,,
\eeqa
and $\Omega_x$ and $w_x$ given by Eqs.~(\ref{Omex}) 
and (\ref{wx}) with $g_{,Y_1}=g_{,Y_2}=1$ and $w_m=0$.

\subsection{Kinetic fixed points}

The kinetic fixed points correspond to non-zero values 
of $x_i$ ($i=1,2,\cdots,N$) with
\be
y=0\,.
\ee
Since the variables $Y_i=x_i^2/y^2$ diverge in this case, 
this restricts the Lagrangian (\ref{scalinglag}) to 
specific forms.
Since the pressure $P$ is proportional to $y^2g$, 
the function $g$ can contain the 
terms linear in $Y_i$, but not the terms 
like $Y_i^n$ ($n \geq 2$). Let us then consider the 
function of the form 
\be
g= \sum_{i=1}^{N} c_iY_i
+\sum_{i=1}^{N} \sum_{n>0} c_{-n}^{(i)} Y_i^{-n}
-h(Z_1,\cdots,Z_{N-1})\,,
\label{gchoice}
\ee
where $c_i$ and $c_{-n}^{(i)}$ are constants with positive 
integer $n$, and $h$ is a finite function with 
respect to $Z_1,\cdots,Z_{N-1}$. 
In order to avoid the appearance of ghosts, 
we focus on the case $c_i>0$.
For the choice (\ref{gchoice}) we have
\be
gy^2\,(y \to 0)=\sum_{i=1}^{N}c_i x_i^2\,,\qquad
g_{,Y_i}\,(y \to 0)=c_i\,.
\label{glimit}
\ee
We also consider the case in which $g_{,Z_i}$ is finite, 
such that $y^2 g_{,Z_{,i}} \to 0$ for $y \to 0$.  
Equations (\ref{x1}) and (\ref{x2}) read 
\be
x_i'=\frac32 x_i \left[ w_m+(w_x-w_m)\Omega_x-1 \right]
-\frac{\sqrt{6}}{2c_i} Q_i (1-\Omega_x)\,,\qquad
(i=1,2,\cdots,N).
\label{xieq}
\ee

{}From Eqs.~(\ref{Omew0}), (\ref{Omew}) and (\ref{glimit}), 
we obtain $\Omega_x (w_x-1)=0$, that is, 
$\Omega_x=0$ or $w_x=1$.
When $\Omega_x=0$ we have $\sum_{i=1}^{N}c_ix_i^2=0$ 
and hence $x_i=0$ for $c_i>0$ ($i=1,2,\cdots,N$). 
{}From Eq.~(\ref{xieq}) this can be realized only for $Q_i=0$. 
When $w_m=0$, this corresponds to the standard 
matter-dominated epoch.

In the rest of this section we study the case 
\be
w_x=1\,.
\label{wx1}
\ee
Then, Eq.~(\ref{xieq}) reads
\be
x_i'=\frac12 \left[ 3(w_m-1)x_i
-\frac{\sqrt{6}Q_i}{c_i} \right] (1-\Omega_x) \,.
\label{xieq2}
\ee
There are two qualitatively different fixed points: 
\begin{itemize}
\item{Point C: a $\phi$-matter-dominated era ($\phi$MDE) \cite{Amendola} with}
\beqa
x_i=\frac{\sqrt{6}Q_i}{3c_i (w_m-1)}\,,
\qquad (i=1,2\cdots,N).
\label{xiso}
\eeqa
\item{Point D: a purely kinetic solution with}
\beqa
\Omega_x=1\,.
\eeqa
\end{itemize}

\subsubsection{Point C: the $\phi$MDE}

Substituting the solutions (\ref{xiso}) into 
$\Omega_x=\sum_{i=1}^{N} c_ix_i^2$, we obtain
\be
\Omega_x=\frac{2}{3(w_m-1)^2} 
\sum_{i=1}^{N} \frac{Q_i^2}{c_i}\,.
\label{Omexphi}
\ee
This corresponds to the scaling solution appearing in the 
presence of the couplings $Q_i$. 
The effective equation of state is given by 
\be
w_{\rm eff}=w_m+\frac{2}{3(1-w_m)} 
\sum_{i=1}^{N} \frac{Q_i^2}{c_i}\,.
\ee
When $w_m=0$, it follows that 
$w_{\rm eff}=\Omega_x=(2/3)\sum_{i=1}^{N} Q_i^2/c_i$. 
For $Q_i \neq 0$ the $\phi$MDE can replace the standard 
matter era characterized by $w_{\rm eff}=\Omega_x=0$.
For the consistency with observations we require that 
$w_{\rm eff} \ll 1$, by which the couplings $Q_i$ are 
bounded from above \cite{Pettorino}. 
The $\phi$MDE should be followed by some fixed points 
like A and B to enter the regime of cosmic acceleration.

For the derivation of the $\phi$MDE discussed above
the constancy of the terms $Z_i$ is not actually required, so 
the relation $x_{i}=\lambda_1 x_1/\lambda_{i}$
does not need to hold. 
This property is different from that discussed for the scaling 
solution in Sec.~\ref{autosec}.

\subsubsection{Point D: the purely kinetic solution}

Substituting $\Omega_x=1$ and $w_x=1$ into Eq.~(\ref{Omew}), 
we obtain 
\be
\sum_{i=1}^{N} c_i x_i^2=1\,.
\label{sumD}
\ee
If the variables $Z_i$ are constant, there is the relation 
$x_{i}=\lambda_1 x_1/\lambda_{i}$ and hence $x_1$ 
is known from Eq.~(\ref{sumD}).
However, this condition is not mandatory because 
$Z_i$ can vary in time.
{}From Eqs.~(\ref{weff2}) and (\ref{wx1}) we have
\be
w_{\rm eff}=w_x=1\,.
\ee
This solution can be used neither for the matter era 
nor for the epoch of cosmic acceleration.

The four fixed points are summarized in Table \ref{table1}. 

\begin{table*}[t]
\begin{center}
\begin{tabular}{|c|c|c|c|c|}
\hline
 &  $x_1$ & $\Omega_x$ & $w_{\rm eff}$ & $w_x$ \\
\hline
\hline
A & $\frac{\lambda^2}{\sqrt{6}\lambda_1}$ & 1 & 
$-1+\frac{\lambda^2}{3}$ & $-1+\frac{\lambda^2}{3}$ \\
\hline
B & $\frac{\sqrt{6}(1+w_m)}{2\lambda_1(1+q)}$  
&  $\frac{3(1+w_m)+q(1+q)\lambda^2}{(1+q)^2\lambda^2}$ & 
$\frac{w_m-q}{1+q}$ &
$\frac{3w_m(1+w_m)-q(1+q)\lambda^2}{3(1+w_m)+q(1+q)\lambda^2}$ \\
\hline
C &  $\frac{\sqrt{6}Q_1}{3c_1(w_m-1)}$ & 
$\frac{2}{3(w_m-1)^2}\sum_{i=1}^N\frac{Q_i^2}{c_i}$ & 
$w_m+\frac{2}{3(1-w_m)} 
\sum_{i=1}^{N} \frac{Q_i^2}{c_i}$ & $1$ \\
\hline
D & $\sum_{i=1}^Nc_ix_i^2=1$  & 1 & 
$1$ & $1$ \\
\hline
\end{tabular}
\end{center}
\caption[crit]{The properties of four fixed points A, B, C, and D, where 
$\lambda$, $q$ and $c_i$ are defined in Eqs.~(\ref{lamdef}), 
(\ref{qdef}) and (\ref{gchoice}), respectively. 
The points A and B exist for the general $N$-field Lagrangian 
(\ref{scalinglag}), whereas the points C and D arise for the specific 
Lagrangian $P=e^{-\lambda_1 \phi_1}g$ with 
$g$ given by Eq.~(\ref{gchoice}).
}
\label{table1} 
\end{table*}

\section{Stability of the fixed points A and B}
\label{stasec}

In this section we study the stability of the fixed points A and B 
in the presence of a barotropic perfect fluid with the equation of state $w_m$.
We consider the two-field model characterized by the Lagrangian
\be
P=e^{-\lambda_1 \phi_1} g(Y_1,Y_2,Z_1)\,.
\label{lagtwo}
\ee
For the point A it is possible to carry out the general 
analysis without restricting the functional form of $g(Y_1,Y_2, Z_1)$.
Since the general stability analysis for the point B is very complicated, 
we focus on a simpler model with two canonical fields.
We also interpret the stability conditions of the scaling solution B 
as a geometric perspective in the two-field space.
For the stability of the other fixed points like C and D, 
we will discuss it in Sec.~\ref{example} for a more 
concrete two-field model.

\subsection{Fixed point A}
\label{pointA}

Let us first study the stability of the fixed point A characterized 
by the conditions (\ref{sca1})-(\ref{weffA}).
Taking the ${\cal N}$-derivative of the field density parameter 
$\Omega_x=2(x_1^2g_{,Y_1}+x_2^2g_{,Y_2})-gy^2$
and using Eqs.~(\ref{x1})-(\ref{y}), we obtain
\be
\Omega_x'=(\Omega_x-1)\left[ 3(w_x-w_m)\Omega_x
+\sqrt{6} (Q_1x_1+Q_2 x_2) \right]\,.
\ee
Perturbing this equation at linear order and denoting the 
perturbed quantities like $\delta \Omega_x$, we have
\be
\delta \Omega_x'=
\left[ 3(w_x-w_m)\Omega_x
+\sqrt{6} (Q_1x_1+Q_2 x_2) \right] \delta \Omega_x
+(\Omega_x-1) \left[3y^2 \delta g+6gy \delta y
-3w_m \delta \Omega_x+\sqrt{6}
(Q_1 \delta x_1+Q_2 \delta x_2) \right]\,.
\label{delOme}
\ee
Since $\Omega_x=1$, $w_x=-1+\lambda^2/3$, and 
$x_i=\lambda^2/(\sqrt{6}\lambda_i)$ for the point A, 
it follows that 
\be
\delta \Omega_x'=
\left[ \lambda^2 \left( 1+q \right)-3(1+w_m) \right]
\delta \Omega_x\,.
\ee
Hence the point A is stable in the $\Omega_x$ direction 
under the condition 
\be
\lambda^2(1+q)<3(1+w_m)\,.
\label{Acon1}
\ee
Once this condition is satisfied, then we need to study the behavior 
of the other variables. The presence of $\delta\Omega_x$ acts as 
an explicitly time-dependent inhomogeneous 
term in the other perturbed equations of motion.
Therefore, in order to study the stability of the other variables, 
we will consider only the homogeneous contribution, and we will set 
$\delta\Omega_x$ to vanish in the remaining equations.

Let us consider the quantities $Y_1$, $Y_2$, and $Z_1$ as other dynamical 
variables. Then we can express both $\delta Y_1$ and $\delta Y_2$ 
in terms of $\delta Z_1$ and its derivatives. 
Finally, we obtain a third-order differential equation for $\delta Z_1$. 
On looking for a solution of the kind 
$\delta Z_1\propto e^{\Gamma {\cal N}}$, 
we find that $\Gamma$ satisfies a cubic equation, 
which can be factorized as
\begin{equation}
(2\Gamma+6-\lambda^2)(\alpha\Gamma^2+\beta\Gamma+\delta)=0\, ,
\label{eq:cubic}
\end{equation}
where
\begin{align}
\alpha &=2 \lambda _1^2 \lambda _2^2 Y_1 \left[
2 \lambda _1^4\lambda^2 Y_1 (g_{,Y_1} g_{,Y_2Y_2}-2 Y_1 g_{,Y_1Y_2}^2
+2 Y_1 g_{,Y_1Y_1} g_{,Y_2Y_2})+\lambda _1^2 \lambda_2^2 \lambda^2
g_{,Y_2}(g_{,Y_1}+2 Y_1 g_{,Y_1Y_1}) \right]\,,\\
\beta &=\frac12\,(6-\lambda^2)\,\alpha\,,\\
\delta &=\lambda _2^2 \Bigl[ Y_1 \lambda ^2 
\lambda _1^2 g_{,Y_1 Y_2}\left\{Y_1 \lambda _2 \left[2 \lambda ^2 
\left(\lambda ^2-6\right) \lambda _1^2 g_{,Y_1 Z_1}-\lambda_2 
\left(\lambda ^2 g_{,Y_1}-\lambda _1^2\right) \left(12 \lambda ^2 g_{,Y_1}
+\left(\lambda^2-18\right) \lambda _1^2\right)\right]-4 \lambda ^4 
\lambda _1^2 g_{,Z_1Z_1}\right\} \nonumber\\
&+\lambda _2^2\Bigl\{ 2 Y_1 \lambda _1^2 \lambda ^6 \left(g_{,Y_1 Z_1}^2-
g_{,Y_1Y_1} g_{,Z_1Z_1}\right)
-3 Y_1\left(\lambda ^2-2\right) \lambda _1^2 \lambda _2 \lambda ^2 
g_{,Y_1 Z_1} \left(\lambda ^2g_{,Y_1}-\lambda _1^2\right)\nonumber\\
&+Y_1 \lambda _2^2 \left(\lambda _1^2-\lambda ^2 g_{,Y_1}\right){}^2
   \left[\left(\lambda ^2-6\right) \lambda _1^2-6 Y_1 \lambda ^2 g_{,Y_1Y_1}\right]
-\lambda _1^4 \lambda
   ^4 g_{,Z_1Z_1}\Bigr\}\Bigr]+ 2 Y_1 \lambda ^6 \lambda _1^6 g_{,Y_2 Z_1}^2
   \nonumber\\
 &+Y_1 \lambda ^4 \lambda _2 
\lambda_1^4 g_{,Y_2 Z_1} \left[4
   \lambda _2 \lambda ^2 g_{,Y_1 Z_1}-2 Y_1 \left(\lambda ^2-6\right) 
\lambda _1^2 g_{,Y_1 Y_2}
+\lambda_2^2 \left(2 \lambda_1^2-3 \left(\lambda ^2-2\right) g_{,Y_1}-2 Y_1 
\left(\lambda ^2-6\right) g_{,Y_1Y_1}\right)\right]\nonumber\\
&+Y_1 \lambda ^2 \lambda _1^4 g_{,Y_2Y_2} \left\{ 
Y_1 \lambda _2 
\left[2 \lambda ^2 \left(\lambda ^2-6\right) \lambda _1^2 g_{,Y_1 Z_1}-\lambda_2 
\left(\lambda ^2 g_{,Y_1}-\lambda_1^2\right) \left(6 \lambda ^2 g_{,Y_1}+
\left(\lambda ^2-12\right) \lambda _1^2\right)\right]
-2\lambda ^4 \lambda _1^2 g_{,Z_1Z_1}\right\}\,.
\end{align}

The three solutions to Eq.~(\ref{eq:cubic}) are the following
\beqa
\Gamma_1&=&-\frac12 (6-\lambda^2)\,,\\
\Gamma_{2,3}&=&
-\frac14 (6-\lambda^2) \left[ 1 \pm 
\sqrt{1-\frac{4\alpha \delta}{\beta^2}}
\right]\,.
\eeqa
The stability of the point A demands that $\Gamma_{1}$, $\Gamma_2$, 
$\Gamma_3$ are negative, or if they are imaginary, they have negative 
real parts. This requires the following conditions
\beqa
& & \lambda^2<6\,,\label{Acon2} \\
& & \alpha \delta>0\,.
\label{Acon3}
\eeqa
If the fixed point A is responsible for cosmic acceleration, then $w_{\rm eff}=-1+\lambda^2/3<-1/3$ and 
$\lambda^2<2$. In this case the condition (\ref{Acon2}) is automatically satisfied.
In summary, the point A is stable under the conditions 
(\ref{Acon1}), (\ref{Acon2}), and (\ref{Acon3}).

For concreteness, we study the model 
\be
g(Y_1,Y_2,Z_1)=Y_1+Y_2-h(Z_1)\,,
\label{twocanonical}
\ee
in which case the field potential is given by 
Eq.~(\ref{twofieldpo}). For the positivity of the potential 
we demand the condition $h(Z_1)>0$.
{}From Eq.~(\ref{sca1}) we have that 
$gy^2=x_1^2+x_2^2-y^2h=-1+\sqrt{6}\lambda_1 x_1/3$.
Substituting $x_1=\lambda^2/(\sqrt{6}\lambda_1)$ and 
$x_2=\lambda^2/(\sqrt{6}\lambda_2)$ into this relation, 
it follows that 
\be
y^2=\frac{1}{h} \left( 1-\frac{\lambda^2}{6} \right)\,,\qquad
Y_1=\frac{\lambda^4}{\lambda_1^2 (6-\lambda^2)}h\,.
\label{yA}
\ee
Combining Eq.~(\ref{sca3}) with the first of Eq.~(\ref{yA}), we obtain
\be
h_{,Z_1}=-\frac{\lambda^2}{\lambda_2}h\,.
\label{hZre0}
\ee
The quantities $\alpha$ and $\delta$ read
\be
\alpha=\frac{2\lambda_1^2 \lambda_2^4 \lambda^6}
{6-\lambda^2}h\,,\qquad
\delta=(\lambda_1 \lambda_2 \lambda)^4
\left( h_{,Z_1 Z_1}-\frac{\lambda^4}{\lambda_2^2}h \right)\,.
\ee
Since $\alpha>0$, the condition 
(\ref{Acon3}) translates to $\delta>0$, that is
\be
h_{,Z_1 Z_1}>\frac{\lambda^4}{\lambda_2^2}h\,.
\label{Acon3d}
\ee
The stability of the fixed point A with cosmic acceleration ($\lambda^2<2$)
is ensured under the conditions (\ref{Acon1}) and (\ref{Acon3d}) 
together with the additional constraint (\ref{hZre0}).

As an example, let us consider the model (\ref{Dodelson}), i.e.,
\be
h(Z_1)=V_1 e^{\mu_1 Z_1}+V_2 e^{\mu_2 Z_1}\,.
\label{hchoice}
\ee
In this case, Eqs.~(\ref{hZre0}) and (\ref{Acon3d}) 
translate to 
\beqa
& &
\frac{V_2}{V_1}=-e^{(\mu_1-\mu_2)Z_1} 
\frac{\mu_1+\lambda^2/ \lambda_2}
{\mu_2+\lambda^2/ \lambda_2}\,,\label{ca1} \\
& &
V_1 \left( \mu_1+\lambda^2/\lambda_2 \right)
\left( \mu_1-\mu_2 \right)>0\,,
\label{ca2}
\eeqa
respectively.
The positivity of the potential ($V_1>0$ and $V_2>0$) 
in Eq.~(\ref{ca1}) requires that  
\be
\left( \mu_1 +\lambda^2/\lambda_2 \right)
\left( \mu_2 +\lambda^2/\lambda_2 \right)<0\,,
\label{mucon}
\ee
under which (\ref{ca2}) is automatically satisfied.
The condition (\ref{mucon}), together with (\ref{Acon1}), 
ensures the stability of the fixed point A.

\subsection{Fixed point B}

We can proceed to study the stability of the point B along the same lines shown 
in Sec.~\ref{pointA}. In this case, however, the equation of motion for 
the variable $\delta\Omega_x$ 
does not decouple from the other variables. This makes the analysis more cumbersome. 
Nonetheless, since we are dealing with linear differential equations with constant coefficients, 
the algorithm for the solution is straightforward. Namely, we can look for solutions of the kind 
$\delta Y_1=A_1\exp(\Gamma {\cal N})$, 
$\delta Y_2=A_2\exp(\Gamma {\cal N})$, 
$\delta Z_1=A_3\exp(\Gamma {\cal N})$,
and $\delta\Omega_x=A_4\exp(\Gamma {\cal N})$. 
Then we can use three of the four differential equations to write e.g.,\ $A_{1,2,3}$ in terms
of $A_3$ and $\Gamma$. The last available equation transforms into
a quartic algebraic equation for the variable $\Gamma$. 
Although the general solutions can be given, 
they are too complicated to be written here.

The analysis can be simplified by specifying the models.
In the following, let us focus on the model 
\be
g(Y_1,Y_2,Z_1)=Y_1+Y_2-h(Z_1)\,,\qquad
{\rm with} \qquad
Q_i=0 \quad (i=1,2)\,.
\label{twocanonical2}
\ee
We perturb the autonomous equations of motion about the scaling 
solution B characterized by Eqs.~(\ref{weffscaling})-(\ref{wx}). 
Since the field density parameter in Eq.~(\ref{Omex}) is given by 
$\Omega_x=x_1^2+x_2^2+hy^2$ with $x_1=\sqrt{6}(1+w_m)/(2\lambda_1)$
and  $x_2=\sqrt{6}(1+w_m)/(2\lambda_2)$, we obtain 
\be
y^2=\frac{3(1-w_m^2)}{2\lambda^2} \frac{1}{h}\,.
\ee
{}From Eq.~(\ref{yg}) there is also the following relation
\be
h_{,Z_1}=-\frac{\lambda^2}{\lambda_2}h\,.
\label{hZre}
\ee

The linear perturbations $\delta x_1$, $\delta x_2$, $\delta y$, and 
$\delta Z_1$ about the scaling fixed point B 
obey the equations of motion 
\be
{}^t (\delta x_1', \delta x_2', \delta y', \delta Z_1')={\cal M}\,\,
{}^t (\delta x_1, \delta x_2, \delta y, \delta Z_1)\,,
\ee
where ${\cal M}$ is a $4 \times 4$ matrix.
The four eigenvalues of the matrix ${\cal M}$ are given by 
\beqa
\gamma_{1,2} &=&
-\frac34 (1-w_m) \left[ 1 \pm \sqrt{1-
\frac{8(1+w_m)[\lambda^2-3(1+w_m)]}{\lambda^2 (1-w_m)}} 
\right]\,,\\
\gamma_{3,4} &=& -\frac34 (1-w_m) \left[ 1 \pm 
\sqrt{1-\frac{8\lambda_1^2(1+w_m)}{\lambda^4 (1-w_m)h}
 \left(h_{,Z_1Z_1}-\frac{\lambda^4}{\lambda_2^2}h \right)}
 \right]\,.
\eeqa
For $0 \le w_m<1$ the scaling solution B is stable provided 
that $\gamma_{1,2,3,4}$ 
are negative or have negative real parts, that is
\beqa
& &\lambda^2 > 3(1+w_m)\,,
\label{Bsta1} \\
& & h_{,Z_1Z_1}> 
\frac{\lambda^4}{\lambda_2^2}h\,.
\label{Bsta2} 
\eeqa
The conditions (\ref{Bsta2}) and (\ref{hZre}) are exactly 
the same as Eqs.~(\ref{Acon3d}) and (\ref{hZre0}) derived for the stability 
of the fixed point A with cosmic acceleration. 
The difference appears for another condition (\ref{Bsta1}), whose 
inequality is opposite to the stability condition (\ref{Acon1}) for 
the point A with $q=0$. This means that, as far as the model (\ref{twocanonical2}) 
is concerned, the scaling solution B is stable (unstable) when the fixed point A 
with cosmic acceleration is unstable (stable).

It is instructive to understand the meaning of the stability conditions
(\ref{Bsta1}) and (\ref{Bsta2}) as well as the additional constraint 
(\ref{hZre}). In doing so, we perform the field rotation analogous to (\ref{rotation}):
\beqa
\phi_1 &=& (\cos \theta) \sigma -(\sin \theta)s\,,\\
\phi_2 &=& (\sin \theta) \sigma+(\cos \theta)s\,,
\eeqa
where $\theta$ is a constant satisfying $\cos \theta=\lambda/\lambda_1$ 
and $\sin \theta=\lambda/\lambda_2$. 
Then the potential (\ref{twofieldpo}) transforms as 
\be
V(\sigma,s)=e^{-\lambda \sigma+(\lambda^2/\lambda_2)Z_1}\,
h(Z_1)\,,\quad {\rm where} \quad
Z_1=\frac{\lambda_1}{\lambda}s\,.
\label{twopo2}
\ee
The field $\sigma$ is transverse to the scaling solution (\ref{phiso}), 
whereas the field $s$, or equivalently to $Z_1$, points to a direction 
orthogonal to $\sigma$. 
The scaling solution satisfies the condition $V_{,s}=0$, that is 
\be
h_{,Z_1}=-\frac{\lambda^2}{\lambda_2}\,h\,,
\label{canocon1}
\ee
which is equivalent to (\ref{hZre}).
On using this condition, the second derivative of $V$ 
with respect to $s$ reads
\be
V_{,ss}=\frac{\lambda_1^2}{\lambda^2}
e^{-\lambda \sigma+(\lambda^2/\lambda_2)Z_1} 
\left( h_{,Z_1 Z_1}-\frac{\lambda^4}{\lambda_2^2} h
\right)\,.
\ee
The stability along the direction of the field $s$
demands that $V_{,ss}>0$, i.e., 
\be
h_{,Z_1 Z_1}>\frac{\lambda^4}{\lambda_2^2} h\,,
\label{canocon2}
\ee
which matches with the condition (\ref{Bsta2}).
Now the scaling solution can be described by 
an effective single field $\sigma$ having a stability to 
the orthogonal direction. 
{}From Eq.~(\ref{twopo2}) the potential is proportional to 
the exponential term $e^{-\lambda \sigma}$ along the $\sigma$ 
direction. Hence the stability of the fixed point B in the presence 
of a barotropic perfect fluid is the same as 
that derived in Refs.~\cite{clw,Tsujikawa06} for the single-field 
exponential potential, that is,
\be
\lambda^2>3(1+w_m)\,,
\label{lamcon}
\ee
which is the same as another stability condition (\ref{Bsta1}).
The above discussion explains the physical meaning of 
the stability conditions (\ref{Bsta1}) and (\ref{Bsta2}).

If we choose the function $h(Z_1)$ of the form (\ref{hchoice}), 
Eqs.~(\ref{canocon1}) and (\ref{canocon2}) give rise to 
the same conditions as Eqs.~(\ref{ca1}) and (\ref{ca2}), respectively.

\section{Concrete model}
\label{example}

In this section we study the cosmology of the two-field model 
described by the Lagrangian (\ref{Plag}) with (\ref{alvalues}), that is 
\be
P=X_1+X_2-V_1e^{-\lambda_1 (1+\mu_1/\lambda_2)\phi_1
+\mu_1 \phi_2}-V_2e^{-\lambda_1 (1+\mu_2/\lambda_2)\phi_1
+\mu_2 \phi_2}\,.
\label{conlag}
\ee
This corresponds to the Lagrangian 
$P=e^{-\lambda_1 \phi_1}g(Y_1,Y_2,Z_1)$ with 
the $g$ given by Eq.~(\ref{Dodelson}).
In addition to non-relativistic matter with the equation of state $w_m=0$, 
we take into account the radiation with the energy density $\rho_r$ 
satisfying the continuity equation $\dot{\rho}_r+4H \rho_r=0$. 
Instead of $Z_1$ and $y$ defined in Eqs.~(\ref{Zi}) and (\ref{ydef}), 
we introduce the following quantities
\be
y_1 \equiv \sqrt{V_1}\, e^{\mu_1 Z_1/2}y\,,
\qquad
y_2 \equiv \sqrt{V_2}\, e^{\mu_2 Z_1/2}y\,,\label{y2def}
\ee
whose squares can be written as 
\be
y_1^2=\frac{V_1}{3H^2}
e^{-\lambda_1(1+\mu_1/\lambda_2)\phi_1+\mu_1\phi_2}\,,
\qquad
y_2^2=\frac{V_2}{3H^2}
e^{-\lambda_1(1+\mu_2/\lambda_2)\phi_1+\mu_2\phi_2}\,.
\ee
These quantities are related to the potential energies 
of the third and fourth terms on the right-hand side (rhs) 
of Eq.~(\ref{conlag}) respectively, 
whereas $x_1^2$ and $x_2^2$ are 
associated with the kinetic energies of the fields $\phi_1$
and $\phi_2$ respectively.
Several quantities appearing in Eqs.~(\ref{x1})-(\ref{z1}) 
can be expressed as 
\be
\Omega_x=x_1^2+x_2^2+y_1^2+y_2^2\,,\qquad
\Omega_x w_x=
x_1^2+x_2^2-y_1^2-y_2^2\,,
\ee
and $g_{,Z_1}y^2=-\mu_1 y_1^2-\mu_2 y_2^2$.

Defining the density parameter of the radiation as 
$\Omega_r=\rho_r/(3H^2)$, we obtain the following 
autonomous equations of motion
\beqa
x_1' &=& \frac12 x_1 \left( 3\Omega_x w_x-3+\Omega_r \right)
+\frac{\sqrt{6}}{2}\lambda_1 \left[ y_1^2+y_2^2
+\frac{1}{\lambda_2} (\mu_1 y_1^2+\mu_2 y_2^2) \right]
-\frac{\sqrt{6}}{2}Q_1 \Omega_m \,,
\label{x1d}\\
x_2' &=& \frac12 x_2 \left( 3\Omega_x w_x-3+\Omega_r \right)
-\frac{\sqrt{6}}{2}(\mu_1 y_1^2+\mu_2 y_2^2)
-\frac{\sqrt{6}}{2}Q_2 \Omega_m\,,
\label{x2d}\\
y_1' &=& \frac{y_1}{2} \left[ \sqrt{6}\mu_1 
\left( x_2-\frac{\lambda_1}{\lambda_2}x_1 \right)
+3\Omega_x w_x+3+\Omega_r
-\sqrt{6} \lambda_1 x_1 \right]\,,
\label{y1d}\\
y_2' &=& \frac{y_2}{2} \left[ \sqrt{6}\mu_2
\left( x_2-\frac{\lambda_1}{\lambda_2}x_1 \right)
+3\Omega_x w_x+3+\Omega_r
-\sqrt{6} \lambda_1 x_1 \right]\,,
\label{y2d}\\
\Omega_r' &=& \Omega_r \left( 3\Omega_x w_x
-1+\Omega_r \right)\,,
\label{Omerd}
\eeqa
where the density parameter of non-relativistic matter is given by 
\be
\Omega_m=1-\Omega_x-\Omega_r\,.
\ee
The effective equation of state (\ref{weff}) reads
\be
w_{\rm eff}=\Omega_xw_x+\frac13 \Omega_r\,.
\ee
In the following we first study the cosmology with $Q_1=0$ and 
$Q_2=0$ (Secs.~\ref{example1} and \ref{example2}) and 
then proceed to the case in which the couplings are 
present (Sec.~\ref{example3}).

\subsection{Scaling fixed points B during the radiation and matter eras}
\label{example1}

Let us first discuss the cosmology driven by the scaling solution B
in the absence of the couplings $Q_i$.
The scaling fixed point B corresponds to the one at which 
both the potential energies of the third and fourth terms 
in Eq.~(\ref{conlag}) provide non-vanishing contributions 
to $\Omega_x$, such that $y_1 \neq 0$ and $y_2 \neq 0$. 
{}From Eq.~(\ref{Omerd}) there are two qualitatively different 
fixed points: (i) $\Omega_r=1-3\Omega_x w_x$ 
and (ii) $\Omega_r=0$. In each case we obtain the following scaling 
fixed points B1 and B2 relevant to the cosmological dynamics 
during the radiation and matter eras, respectively.

\begin{itemize}
\item (i) Point B1
\be
x_1=\frac{2\sqrt{6}}{3\lambda_1}\,,\qquad 
x_2=\frac{2\sqrt{6}}{3\lambda_2}\,,\qquad
y_1^2=\frac{4}{3(\mu_2-\mu_1)\lambda^2} 
\left( \mu_2+\frac{\lambda^2}{\lambda_2}
\right)\,,\qquad
y_2^2=\frac{4}{3(\mu_1-\mu_2)\lambda^2} 
\left( \mu_1+\frac{\lambda^2}{\lambda_2}
\right)\,,
\ee
where $1/\lambda^2=1/\lambda_1^2+1/\lambda_2^2$. 
At this point we have 
\be
\Omega_x=\frac{4}{\lambda^2}\,,\qquad
\Omega_r=1-\frac{4}{\lambda^2}\,,\qquad
\Omega_m=0\,,\qquad
w_{\rm eff}=w_x=\frac13\,.
\ee
\item (ii) Point B2
\be
x_1=\frac{\sqrt{6}}{2\lambda_1}\,,\qquad 
x_2=\frac{\sqrt{6}}{2\lambda_2}\,,\qquad
y_1^2=\frac{3}{2(\mu_2-\mu_1)\lambda^2} 
\left( \mu_2+\frac{\lambda^2}{\lambda_2}
\right)\,,\qquad
y_2^2=\frac{3}{2(\mu_1-\mu_2)\lambda^2} 
\left( \mu_1+\frac{\lambda^2}{\lambda_2}
\right)\,.
\ee
At this point we have 
\be
\Omega_x=\frac{3}{\lambda^2}\,,\qquad
\Omega_r=0\,,\qquad
\Omega_m=1-\frac{3}{\lambda^2}\,,\qquad
w_{\rm eff}=w_x=0\,.
\ee
\end{itemize}

\begin{figure}
\includegraphics[height=3.2in,width=3.3in]{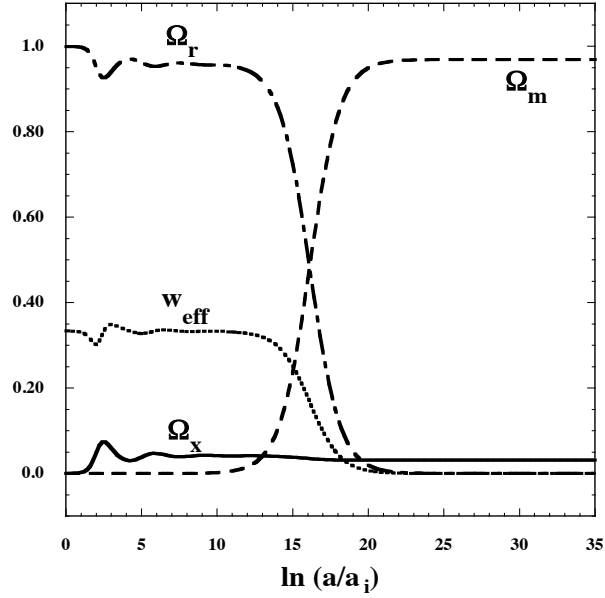}
\caption{\label{fig1}
Evolution of $\Omega_{x}$, $\Omega_{r}$, $\Omega_m$, and $w_{\rm eff}$ 
versus $\ln (a/a_i)$ (where $a_i$ is the initial value of $a$) 
for the model parameters $\lambda_1=13$, 
$\lambda_2=15$, $\mu_1=-3$, $\mu_2=-8$, $Q_1=0$, and 
$Q_2=0$. The initial conditions are chosen to be 
$x_1=0.01$, $x_2=0.02$, $y_1=0.005$, $y_2=0.005$, and 
$\Omega_m=1.0 \times 10^{-7}$. }
\end{figure}
The existence of the above fixed points requires that $y_1^2>0$ and 
$y_2^2>0$, i.e.,
\be
\frac{1}{\mu_2-\mu_1}
\left( \mu_2+\frac{\lambda^2}{\lambda_2}
\right)>0 \qquad {\rm and} \qquad
\frac{1}{\mu_1-\mu_2}
\left( \mu_1+\frac{\lambda^2}{\lambda_2}
\right)>0\,.
\label{conex}
\ee
Hence the signs of the terms $\mu_2+\lambda^2/\lambda_2$ and 
$\mu_1+\lambda^2/\lambda_2$ need to be opposite.

Perturbing Eqs.~(\ref{x1d})-(\ref{Omerd}) about the point B1, 
we find that one of the eigenvalues is 1 and hence 
it is not stable. For the matter scaling point B2 we obtain 
the following five eigenvalues 
\beqa
& &\gamma_{1}=-1\,,
\qquad 
\gamma_{2,3}=-\frac34 \left[ 1 \pm \sqrt{\frac{24}
{\lambda^2}-7} \right]\,,\nonumber \\
& &\gamma_{4,5}=-\frac34 \left[ 1
\pm \sqrt{1+\frac{8}{\lambda_1^2 \lambda_2^4}
\left[\mu_1 (\lambda_1^2+\lambda_2^2)+\lambda_1^2 \lambda_2
\right] \left[ \mu_2 (\lambda_1^2+\lambda_2^2)
+\lambda_1^2 \lambda_2 \right]} \right]\,.
\label{gam1}
\eeqa
Provided $\lambda^2>3$, the eigenvalues $\gamma_{2,3}$ are 
either negative or complex with negative real parts. 
The condition $\lambda^2>3$ is equivalent to (\ref{Bsta1}) 
with $w_m=0$.
The eigenvalues $\gamma_{4,5}$ do not have positive 
real values for
\be
\left( \mu_1+\mu_c \right) \left( \mu_2+\mu_c \right)<0\,,
\qquad
{\rm where} \qquad
\mu_c \equiv \frac{\lambda_1^2 \lambda_2}
{\lambda_1^2+\lambda_2^2}=\frac{\lambda^2}{\lambda_2}\,,
\label{scalingcon}
\ee
which coincides with (\ref{mucon}).
As long as the conditions (\ref{conex}) hold,
the condition (\ref{scalingcon}) is satisfied. 

For the existence of the scaling point B1 during the radiation era,
we require that $\Omega_x<1$ and hence $\lambda^2>4$.
Since the field density parameter is constrained to be 
$\Omega_x<0.045$ from the BBN \cite{Bean}, 
this puts the bound $\lambda>9.4$. 
In this case the fixed point B2 is stable 
under the condition (\ref{scalingcon}).
When the conditions $\lambda^2>4$ and (\ref{scalingcon}) 
are satisfied, the eigenvalues $\gamma_{2,3}$ and $\gamma_{4,5}$
corresponding to the radiation fixed point B1 are negative (or complex
with negative real parts) with $\gamma_1=1$, so it is a saddle point followed 
by the scaling matter solution B2.

In Fig.~\ref{fig1} we plot the evolution of the density parameters
as well as the effective equation of state for $\lambda_1=13$, 
$\lambda_2=15$, $\mu_1=-3$, $\mu_2=-8$, 
$Q_1=0$, and $Q_2=0$. Since $\lambda=9.82$ and 
$\mu_c=6.43$ in this case,  
the stability condition (\ref{scalingcon}) is satisfied.
The initial conditions are chosen such that $y_1$ and $y_2$ as well as
$x_1$ and $x_2$ are the similar orders to each other.
As we see in Fig.~\ref{fig1}, the solution temporally approaches the scaling 
radiation point B1 with $\Omega_x=4/\lambda^2=4.14 \times 10^{-2}$ and 
$w_{\rm eff}=1/3$.
This is followed by the stable scaling matter point B2 characterized by 
$\Omega_x=3/\lambda^2=3.11 \times 10^{-2}$ and 
$w_{\rm eff}=0$. During the matter era, at least 
one of the eigenvalues for the $5 \times 5$ matrix associated 
with the perturbations about the fixed 
points  A, C, and D shown in Table \ref{table2} are positive for the 
model parameters used in Fig.~\ref{fig1}, so they are unstable. 
If $\mu_1$ and $\mu_2$ are outside the range (\ref{scalingcon}), 
we also numerically confirmed that the fixed point B2 
is not stable. 

The above results show that, for the parameters $\lambda_1 \gtrsim 10$, 
$\lambda_2 \gtrsim 10$, and $\mu_1$, $\mu_2$ satisfying the 
condition (\ref{scalingcon}), the solutions enter 
the scaling regime in which the field energy densities 
track the background energy density.
This is a nice feature because even with initially large scalar-field energy 
densities the solutions approach the scaling matter fixed point B2 with 
$\Omega_x/\Omega_m={\rm constant} <O(1)$.
However, if we try to explain the late-time acceleration of the Universe 
as well, we need to consider different parameter spaces and initial conditions 
as compared to those given above. 
In Sec.~\ref{example2} we explore such a possibility 
for the model (\ref{conlag}).

\begin{table*}[t]
\begin{center}
\begin{tabular}{|c|c|c|c|c|c|c|c|}
\hline
 &  $x_1$ & $x_2$ & $y_1^2$ & $y_2^2$ 
& $\Omega_x$ & $\Omega_r$ & $w_{\rm eff}$ \\
\hline
\hline
A & $\frac{\lambda^2}{\sqrt{6}\lambda_1}$ & 
$\frac{\lambda^2}{\sqrt{6}\lambda_2}$ & 
$\frac{[\lambda_1^2 (\lambda_2^2-6)-6\lambda_2^2]
[\lambda_1^2 (\lambda_2+\mu_2)+\lambda_2^2 \mu_2]}
{6(\mu_1-\mu_2)(\lambda_1^2+\lambda_2^2)^2}$ & 
$\frac{[\lambda_1^2 (\lambda_2^2-6)-6\lambda_2^2]
[\lambda_1^2 (\lambda_2+\mu_1)+\lambda_2^2 \mu_1]}
{6(\mu_2-\mu_1)(\lambda_1^2+\lambda_2^2)^2}$ 
& $1$ & $0$ & $-1+\frac{\lambda^2}{3}$  \\
\hline
B1 & $\frac{2\sqrt{6}}{3\lambda_1}$ & 
$\frac{2\sqrt{6}}{3\lambda_2}$ &
$\frac{4}{3(\mu_2-\mu_1)\lambda^2} 
\left( \mu_2+\frac{\lambda^2}{\lambda_2}
\right)$ &
$\frac{4}{3(\mu_1-\mu_2)\lambda^2} 
\left( \mu_1+\frac{\lambda^2}{\lambda_2}
\right)$ &
$\frac{4}{\lambda^2}$ &
$1-\frac{4}{\lambda^2}$ & 
$\frac13$ \\
\hline
B2 & $\frac{\sqrt{6}}{2\lambda_1}$ & 
$\frac{\sqrt{6}}{2\lambda_2}$ &
$\frac{3}{2(\mu_2-\mu_1)\lambda^2} 
\left( \mu_2+\frac{\lambda^2}{\lambda_2}
\right)$ &
$\frac{3}{2(\mu_1-\mu_2)\lambda^2} 
\left( \mu_1+\frac{\lambda^2}{\lambda_2}
\right)$  &
$\frac{3}{\lambda^2}$ &
$0$ & $0$  \\
\hline
C & $-\frac{\sqrt{6}Q_1}{3}$ & 
$-\frac{\sqrt{6}Q_2}{3}$ & $0$ & $0$ &
$\frac23 (Q_1^2+Q_2^2)$ &
$0$ & $\frac23 (Q_1^2+Q_2^2)$  \\
\hline
D & $x_1$ & 
$\pm \sqrt{1-x_1^2}$ & $0$ & $0$ &
$1$ & $0$ & $1$ \\
\hline
E1 & $\frac{2\sqrt{6}\lambda_1 \lambda_2 (\lambda_2+\mu_1)}
{3\xi}$ & 
$-\frac{2\sqrt{6}\lambda_2^2 \mu_1}{3\xi}$ & 
$\frac{4\lambda_2^2}{3\xi}$ & 
$0$ &
$\frac{4\lambda_2^2}
{\xi}$ & 
$1-\frac{4\lambda_2^2}
{\xi}$ & $\frac13$ \\
\hline
E2 & $\frac{\sqrt{6}\lambda_1 \lambda_2 (\lambda_2+\mu_1)}
{2\xi}$ & 
$-\frac{\sqrt{6}\lambda_2^2 \mu_1}{2\xi}$ & 
$\frac{3\lambda_2^2}{2\xi}$ & 
$0$ &
$\frac{3\lambda_2^2}
{\xi}$ & $0$ & $0$ \\
\hline
F & $\frac{\sqrt{6}\lambda_1 (\lambda_2+\mu_1)}{6\lambda_2}$ & 
$-\frac{\sqrt{6}\mu_1}{6}$ & 
$1-\frac{\xi}{6\lambda_2^2}$ & 
$0$ & $1$ & $0$ & $-1+\frac{\xi}{3\lambda_2^2}$ \\
\hline
\end{tabular}
\end{center}
\caption[crit]{The properties of eight fixed points A, B1, B2, C, D, E1, E2, 
and F appearing in the model (\ref{conlag}) in the presence of non-relativistic 
matter ($w_m=0$) and radiation ($w_m=1/3$).
The quantity $\xi$ is defined by 
$\xi=\lambda_2^2 \mu_1^2+\lambda_1^2 (\lambda_2+\mu_1)^2$.}
\label{table2} 
\end{table*}

%
\subsection{Scaling solutions followed by cosmic acceleration}
\label{example2}

If either $\lambda_1$ or $\lambda_2$ is smaller than the order of 1, 
it is possible for the fixed point A to give rise to the late-time 
cosmic acceleration. 
In this section we study the case in which $\lambda_1 \gtrsim O(10)$ 
and $\lambda_2 \lesssim O(1)$ with $Q_1=0$ and $Q_2=0$.
 
The dark energy dynamics based on the separate 
exponential potentials characterized by the Lagrangian 
$P=X_1+X_2-V_1e^{-\lambda_1 \phi_1}
-V_2e^{-\lambda_2 \phi_2}$ have been studied 
in detail in Refs.~\cite{Guo,Blais,Kim,Ohashi,Bruck}.
In such models, i.e., $\mu_1=0$ and $\mu_2=-\lambda_2$,
the potential $V_2e^{-\lambda_2 \phi_2}$ 
is suppressed relative to the other one $V_1e^{-\lambda_1 \phi_1}$ 
during most of the cosmic expansion history, 
but the former comes out at late times to drive the accelerated expansion.
In other words the quantities $y_2$ and $x_2$ are much smaller than 1 
for the redshift $z=1/a-1 \gg 1$, but they grow to the order of 1 
around today ($a=1$). 
If $\mu_1=0$, then it is clear from Eqs.~(\ref{x2d}) and (\ref{y2d}) that the 
variables $x_2$ and $y_2$ remain to be much smaller than 1 for $z \gg 1$ 
because the field $\phi_2$ does not have a direct coupling to $\phi_1$.

When $\mu_1 \neq 0$, however, the presence of the term 
$-\sqrt{6} \mu_1 y_1^2/2$ in Eq.~(\ref{x2d}) generally gives rise to 
a non-negligible kinetic energy of $\phi_2$ relative to $\phi_1$. 
As for the potential energy, Eq.~(\ref{y2d}) shows that $y_2$ can be much 
smaller than $y_1$ during most of the radiation and matter eras 
for $V_2 \ll V_1$. Compared to the scaling fixed points B1 and B2, 
we search for other types of fixed points E1 (radiation era) and 
E2 (matter era) characterized by $y_2=0$ and non-zero values 
of $x_1$, $x_2$, and $y_1$.
They are given, respectively, by 

\begin{itemize}
\item (i) Point E1 
\be
x_1=\frac{2\sqrt{6}\lambda_1 \lambda_2 (\lambda_2+\mu_1)}
{3\xi}
\,,\qquad 
x_2=-\frac{2\sqrt{6}\lambda_2^2 \mu_1}{3\xi}
\,,\qquad 
y_1^2=\frac{4\lambda_2^2}{3\xi} \,,
\qquad
y_2^2=0\,,
\label{Efixedra}
\ee
where 
\be
\xi \equiv \lambda_2^2 \mu_1^2+\lambda_1^2 (\lambda_2+\mu_1)^2\,.
\label{xidef}
\ee
At this point we have
\be
\Omega_x=\frac{4\lambda_2^2}{\xi}\,,\qquad
\Omega_r=1-\frac{4\lambda_2^2}{\xi}\,,\qquad
\Omega_m=0\,,\qquad
w_{\rm eff}=w_x=\frac13\,.
\label{Erad}
\ee
\item (ii) Point E2 
\be
x_1=\frac{\sqrt{6}\lambda_1 \lambda_2 (\lambda_2+\mu_1)}{2\xi}
\,,\qquad 
x_2=-\frac{\sqrt{6}\lambda_2^2 \mu_1}{2\xi}
\,,\qquad 
y_1^2=\frac{3\lambda_2^2}{2\xi}\,,
\qquad
y_2^2=0\,.
\label{Efixedma}
\ee
At this point we have
\be
\Omega_x=\frac{3\lambda_2^2}{\xi}\,,\qquad
\Omega_r=0\,,\qquad
\Omega_m=1-\frac{3\lambda_2^2}{\xi}\,,\qquad
w_{\rm eff}=w_x=0\,.
\label{Emat}
\ee
\end{itemize}

See Table \ref{table2} for the summary of fixed points relevant to 
the cosmological dynamics discussed in this section.
Note that the condition $y_1^2>0$ is automatically satisfied 
for the points E1 and E2.
In the $\mu_1 \to 0$ limit we have $x_2 \to 0$, so 
the field $\phi_2$ is effectively decoupled from the cosmological 
dynamics during the radiation and matter eras.
If $\mu_1 \neq 0$, then the kinetic energy of $\phi_2$ 
contributes to the dynamics such that 
the field density parameter $\Omega_x$ is modified 
relative to the case $\mu_1=0$.
If the solution is in the scaling regime during the BBN epoch, 
the bound $\Omega_x<0.045$ translates to 
\be
\frac{4\lambda_2^2}
{\lambda_2^2 \mu_1^2+\lambda_1^2 (\lambda_2+\mu_1)^2}
<0.045\,.
\label{BBN2}
\ee
The presence of the $\mu_1$  term allows the possibility of 
reducing the lhs of Eq.~(\ref{BBN2}) relative to the case $\mu_1=0$. 

\begin{figure}
\includegraphics[height=3.3in,width=3.3in]{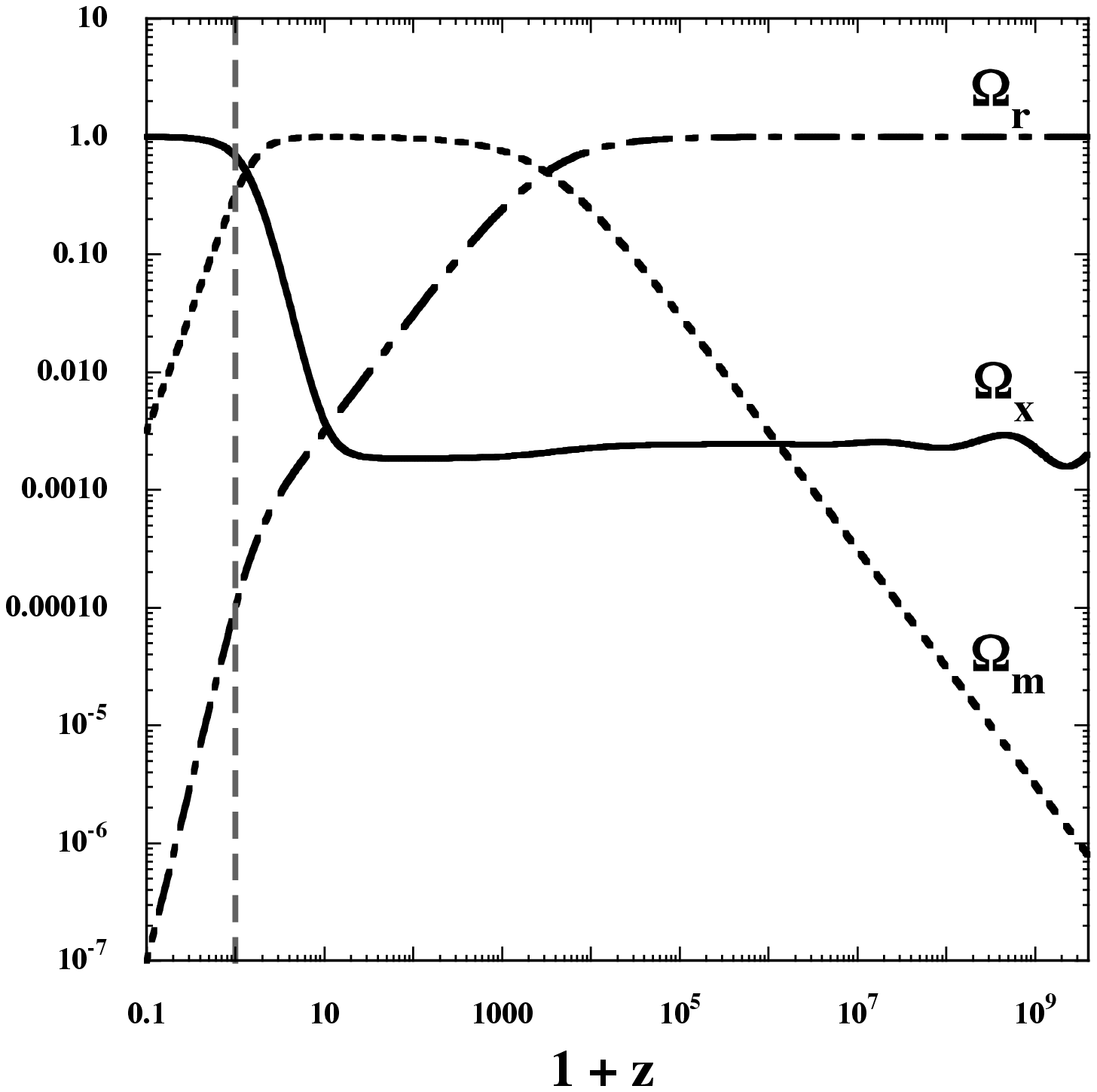}
\includegraphics[height=3.3in,width=3.3in]{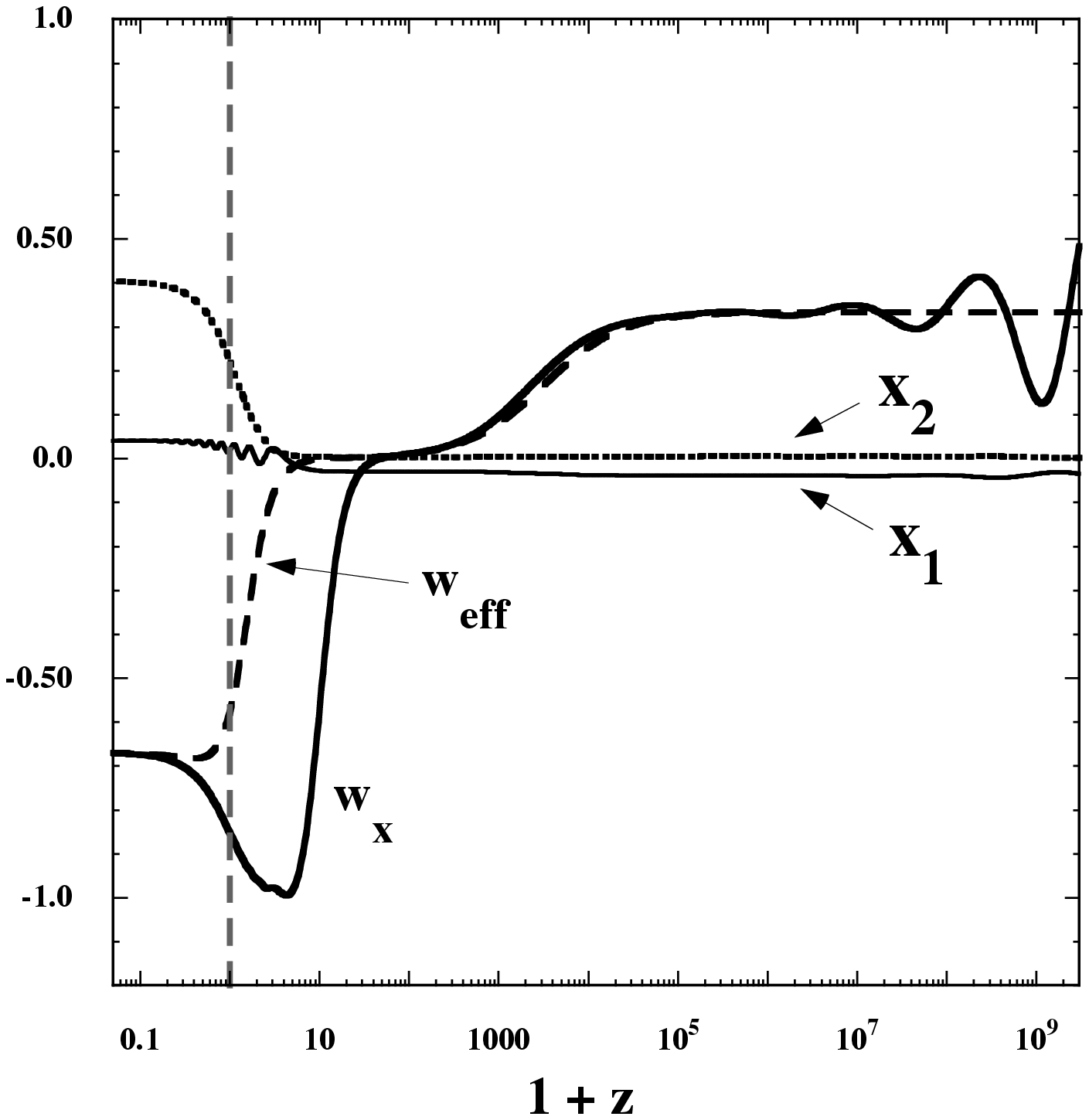}
\caption{\label{fig2}
(Left) Evolution of $\Omega_{x}$, $\Omega_{r}$ and $\Omega_m$ 
versus $1+z$ for the model parameters $\lambda_1=10$, 
$\lambda_2=1$, $\mu_1=-5$, $\mu_2=-1/2$, $Q_1=0$, 
and $Q_2=0$. The initial conditions are chosen to be 
$x_1=-0.04$, $x_2=0$, $y_1=0.02$, $y_2=6.3 \times 10^{-20}$, 
and $\Omega_m=8.0 \times 10^{-7}$. 
(Right) Evolution of $w_{\rm eff}$, $w_x$, $x_1$, and 
$x_2$ for the same model parameters and the initial 
conditions as those in the left panel. 
The vertical dashed line corresponds to the 
present epoch ($z=0$). }
\end{figure}

Perturbing Eqs.~(\ref{x1d})-(\ref{Omerd}) about the matter scaling solution E2, 
the eigenvalues of the matrix associated with the perturbations 
are given by 
\be
\gamma_1=-\frac32\,,\qquad \gamma_2=-1\,,\qquad
\gamma_3=\frac3{2\xi}\,[(\lambda_1^2+\lambda_2^2)\mu_1+
\lambda_1^2 \lambda_2](\mu_1-\mu_2)\,,\qquad
\gamma_{4,5}=-\frac34 \left[ 1 \pm
\sqrt{\frac{24\lambda_2^2}\xi - 7}
\right]\,.
\ee
The point E2 is stable under the conditions
\beqa
& &
[(\lambda_1^2+\lambda_2^2)\mu_1+
\lambda_1^2 \lambda_2]
(\mu_1-\mu_2)<0\,,\label{Esta0} \\
& &
\xi>3\lambda_2^2\,.
\label{Esta}
\eeqa
If $\mu_1=0$ and $\mu_2=-\lambda_2$, for example, 
the condition (\ref{Esta0}) is violated (whereas the condition
(\ref{Esta}) translates to $\lambda_1^2>3$), 
so that the matter point E2 is a saddle.
In order to realize the late-time accelerated expansion, 
either (\ref{Esta0}) or  (\ref{Esta}) should be at least violated 
to exit from the scaling matter era. 
Two of the eigenvalues for the perturbations about the radiation 
point E1 are $1$ and $-1$, so it is a saddle point.

The matter scaling point E2 can be followed 
by the scalar-field dominated solution A discussed in Sec.~\ref{autosec}. 
For the model (\ref{conlag}) the fixed point A 
corresponds to $x_1=\lambda^2/(\sqrt{6}\lambda_1)$, 
$x_2=\lambda^2/(\sqrt{6}\lambda_2)$, and 
\be
y_1^2=\frac{[\lambda_1^2 (\lambda_2^2-6)-6\lambda_2^2]
[\lambda_1^2 (\lambda_2+\mu_2)+\lambda_2^2 \mu_2]}
{6(\mu_1-\mu_2)(\lambda_1^2+\lambda_2^2)^2}\,,\quad
y_2^2=\frac{[\lambda_1^2 (\lambda_2^2-6)-6\lambda_2^2]
[\lambda_1^2 (\lambda_2+\mu_1)+\lambda_2^2 \mu_1]}
{6(\mu_2-\mu_1)(\lambda_1^2+\lambda_2^2)^2}\,.
\quad
\ee
The eigenvalues of the matrix for the perturbations 
about the point A are given by 
\beqa
& &\gamma_1=\lambda^2-4\,,\qquad 
\gamma_2=\lambda^2-3\,,\qquad 
\gamma_3=\frac12( \lambda^2 -6)\,,\nonumber \\
& &\gamma_{4,5}=\frac{\lambda^2-6}{4} 
\left[ 1 \pm \sqrt{1+
\frac{8\lambda^2[(\lambda_1^2+\lambda_2^2)\mu_1 +\lambda_1^2 \lambda_2]
[(\lambda_1^2+\lambda_2^2)\mu_2+\lambda_1^2 \lambda_2]}
{\lambda_1^2 \lambda_2^4 (6-\lambda^2)}} 
\right]\,.
\label{eigenA}
\eeqa
If we impose the condition 
\be
\lambda^2<2\,,
\label{cosace}
\ee
for the realization of cosmic acceleration, we find that the fixed point A 
is stable under the condition 
\be
\left( \mu_1+\mu_c \right) \left( \mu_2+\mu_c \right)<0\,,
\label{scalingcon2}
\ee
which is the same as (\ref{scalingcon}).
The above results are consistent with the general stability 
analysis performed in Sec.~\ref{stasec}.

As long as the condition (\ref{scalingcon2}) holds, the condition (\ref{Esta0}) is 
violated. Hence the matter scaling point E2 
is in fact a saddle that can be followed by the attractor solution A.
When $\lambda$ is in the range (\ref{cosace}), the eigenvalue $\gamma_3$ 
for the matter scaling fixed point B2 in Eq.~(\ref{gam1}) is positive. 
In this case the scaling matter fixed point B2 cannot be a stable attractor.

In Fig.~\ref{fig2} we plot the evolution of the density parameters 
as well as $w_{\rm eff}$, $w_x$, $x_1$, $x_2$ for 
$\lambda_1=10$, $\lambda_2=1$, $\mu_1=-5$, 
$\mu_2=-1/2$ with the initial conditions 
$x_1=-0.04$, $x_2=0$, $y_1=0.02$, 
$y_2=6.3 \times 10^{-20}$, and 
$\Omega_m=8.0 \times 10^{-7}$.
In this case the condition (\ref{Esta0}) does not hold, 
whereas both the conditions (\ref{cosace}) and (\ref{scalingcon2}) are satisfied.
Hence the saddle matter point E2 is followed by the stable attractor A
with cosmic acceleration.

The field density parameter computed from 
Eqs.~(\ref{Erad}) and (\ref{Emat}) are given by 
$\Omega_x=2.46 \times 10^{-3}$ and 
$\Omega_x=1.85 \times 10^{-3}$ during the radiation and 
matter eras, respectively. These values show good 
agreement with the numerical simulation  
of Fig.~\ref{fig2}. We also confirmed that the variables 
$x_1$, $x_2$, and $y_1^2$ first approach the radiation 
scaling solution E1 given by Eq.~(\ref{Efixedra}) and then they evolve to 
the matter scaling solution E2 characterized by Eq.~(\ref{Efixedma}).
From Eq.~(\ref{Efixedma}) we have $x_1=-3.01 \times 10^{-2}$ 
and $x_2=3.77 \times 10^{-3}$ during the matter-dominated epoch 
(see the right panel of Fig.~\ref{fig2}), so the field kinetic 
energy of $\phi_2$ is non-negligible relative to that of $\phi_1$.

The variable $y_2$ is much smaller than 1 for the redshift $z \gg 1$, 
but it grows to the order of 1 for $z<O(1)$.
For the model parameters used in Fig.~\ref{fig2}, 
we have $\lambda=0.995$, 
$x_1=4.04 \times 10^{-2}$, $x_2=0.404$, 
$y_1^2=9.09 \times 10^{-2}$, and $y_2^2=0.744$ 
at the fixed point A.
As we see in the right panel of Fig.~\ref{fig2}, 
the solutions finally approach the fixed point A 
characterized by $w_{\rm eff}=w_x=-1+\lambda^2/3$. 
In the early radiation era the dark energy equation of state 
$w_x$ exhibits an oscillation before reaching the radiation 
scaling point E1 ($w_x=1/3$) and then it evolves to 0 
during the matter era. The decrease of $w_x$ from 0 
occurs around $z=O(10)$ and then $w_x$ reaches a minimum
around $-1$ for $z \gtrsim O(1)$. Finally, $w_x$ approaches 
the asymptotic value $-1+\lambda^2/3$ with $\Omega_x=1$.

The fixed point A corresponds to the assisted inflationary 
attractor with $\lambda$ satisfying 
$1/\lambda^2=1/\lambda_1^2+1/\lambda_2^2$.
Even if each $\lambda_i^2$ is larger than $2$, it is possible to 
realize $\lambda^2<2$ and $w_{\rm eff}=w_x<-1/3$.
The presence of more than two scalar fields allows the possibility of 
reducing $\lambda$ further.

\subsection{$\phi$MDE followed by cosmic acceleration}
\label{example3}

In the presence of the couplings $Q_1$ and $Q_2$, the standard matter 
era can be replaced by the $\phi$MDE. The latter corresponds to 
the case with a negligible potential energy relative to kinetic energies 
of scalar fields, such that $y_1=0$ and $y_2=0$.
The $\phi$MDE is characterized by the fixed point
\be
x_1=-\frac{\sqrt{6}Q_1}{3}\,,\qquad
x_2=-\frac{\sqrt{6}Q_2}{3}\,,\qquad
y_1=0\,,\qquad
y_2=0\,,\qquad \Omega_r=0\, .
\label{phiMDEpo}
\ee
The eigenvalues of the matrix for perturbations about the 
point (\ref{phiMDEpo}) are given by 
\beqa
& &\gamma_1=2 (Q_1^2+Q_2^2)-1\,,\qquad \gamma_{2,3}=
Q_1^2+Q_2^2-\frac32\,,\qquad 
\gamma_4=\frac32+Q_1^2+Q_2 (Q_2-\mu_1)+
\frac{\lambda_1}{\lambda_2} (\mu_1+\lambda_2)Q_1\,,\nonumber \\
& &\gamma_5=\frac32+Q_1^2+Q_2 (Q_2-\mu_2)+
\frac{\lambda_1}{\lambda_2} (\mu_2+\lambda_2)Q_1\,.
\eeqa
In the limit $Q_1 \to 0$ and $Q_2 \to 0$ we have $\gamma_1 \to -1$, 
$\gamma_{2,3} \to -3/2$, and $\gamma_{4,5} \to 3/2$, so the point 
(\ref{phiMDEpo}) is a saddle. 

In the following we shall study the case $\lambda_1 \lesssim O(1)$ and
$\lambda_2 \lesssim O(1)$ with the initial values of $|y_1|$ and 
$|y_2|$ much smaller than 1. 
In this case the potential energy of the scalar fields is much smaller than 
the energy density of the background fluid by the end of the matter era.
If we choose the initial conditions where either $y_1$ or $y_2$ 
is not very close to 0, the solutions can approach other 
fixed points like E1 and E2 discussed in Sec.~\ref{example2}.

\begin{figure}
\includegraphics[height=3.3in,width=3.3in]{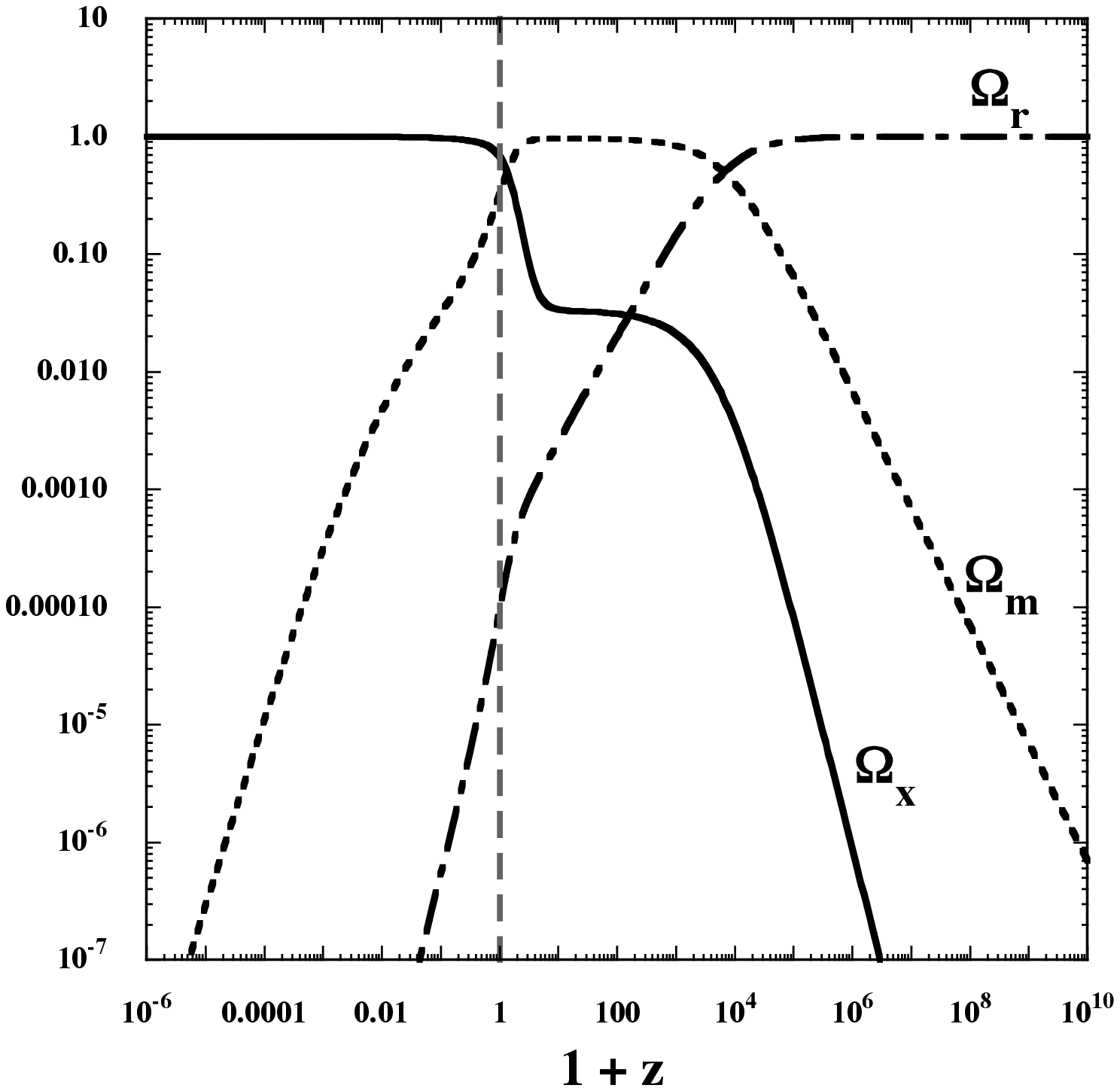}
\includegraphics[height=3.3in,width=3.3in]{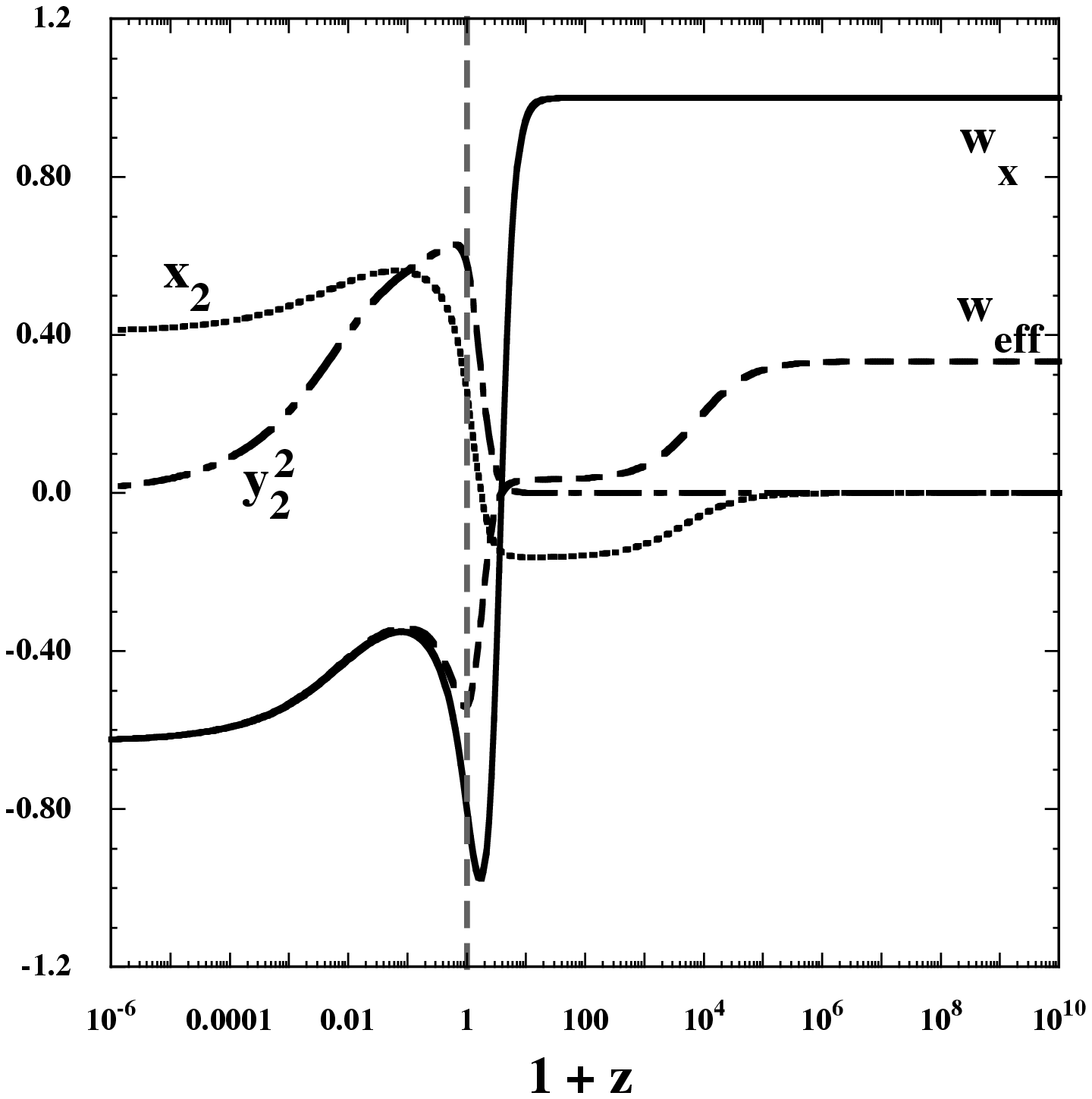}
\caption{\label{fig3}
(Left) Evolution of $\Omega_{x}$, $\Omega_{r}$ and $\Omega_m$ 
versus $1+z$ for the model parameters $\lambda_1=1$, 
$\lambda_2=1.5$, $\mu_1=-1$, $\mu_2=-1.5$, $Q_1=0.1$, 
and $Q_2=0.2$. The initial conditions are chosen to be 
$x_1=-2.0 \times 10^{-8}$, $x_2=-3.0 \times 10^{-8}$,
$y_1^2=1.0 \times 10^{-40}$, $y_2^2=1.0 \times 10^{-39}$, 
and $\Omega_m=4.0 \times 10^{-7}$. 
(Right) Evolution of $w_{\rm eff}$, $w_x$, $x_2$, and 
$y_2^2$ for the same model parameters and the initial 
conditions as those in the left panel. 
The vertical dashed line corresponds to the 
present epoch ($z=0$). }
\end{figure}

In Fig.~\ref{fig3} we plot the evolution of the density parameters 
as well as $w_{\rm eff}$, $w_x$, $x_2$, and $y_2^2$ for $\lambda_1=1$, 
$\lambda_2=1.5$, $\mu_1=-1$, $\mu_2=-1.5$, $Q_1=0.1$, 
and $Q_2=0.2$ with the initial values of $y_1$ and $y_2$ 
close to 0. During the radiation era the field density parameter 
$\Omega_x$ is much smaller than $\Omega_r$, but it temporarily 
approaches the $\phi$MDE characterized by 
\be
\Omega_x=w_{\rm eff}=\frac23 \left( Q_1^2+Q_2^2 \right)\,,
\qquad w_x=1\,.
\ee
These analytic values show good agreement with the numerical simulation 
of Fig.~\ref{fig3} (i.e., $\Omega_x=w_{\rm eff} \simeq 0.033$ with 
$x_2 \simeq -0.163$). 
The dark energy equation of state $w_x$ is close to 1 
during the deep matter era. 
This property can be observationally distinguished from 
the case shown in Fig.~\ref{fig2} (i.e., $w_x \simeq 0$).

In Fig.~\ref{fig3} the solution exits from the $\phi$MDE to the epoch 
of cosmic acceleration. 
In fact, $w_x$ decreases to the minimum close to $-1$ by today 
and then it starts to grow to the asymptotic value around $-0.63$. 
For the model parameters used in Fig.~\ref{fig3} the attractor 
is different from the fixed point A discussed 
in Sec.~\ref{example2}.
In fact one of the eigenvalues in Eq.~(\ref{eigenA}) is
positive ($\gamma_5=0.649$), so the point A is not stable.
In this case the solutions approach another fixed point F 
characterized by 
\be
x_1=\frac{\sqrt{6}\lambda_1 (\lambda_2+\mu_1)}{6\lambda_2}\,,
\qquad
x_2=-\frac{\sqrt{6}\mu_1}{6}\,,
\qquad
y_1^2=1-\frac{\xi}{6\lambda_2^2}\,,\qquad 
y_2^2=0\,, 
 \label{pointE}
\ee
and 
\be
w_{\rm eff}=w_x=-1+\frac{\xi}{3\lambda_2^2}\,,\qquad 
\Omega_x=1\,,\qquad \Omega_r=0\,,
\ee
where $\xi$ is defined by Eq.~(\ref{xidef}).
The eigenvalues of the matrix for perturbations 
about the point F are 
\beqa
& &
\gamma_1=-4+\frac{\xi}{\lambda_2^2}\,,\qquad
\gamma_{2,3}=-3+\frac{\xi}{2\lambda_2^2}\,,\qquad
\gamma_4=
\frac{1}{2\lambda_2^2} \left[ \mu_1 (\lambda_1^2+
\lambda_2^2)+\lambda_1^2 \lambda_2 \right]
(\mu_1-\mu_2)\,, \nonumber \\
& &
\gamma_5=-3+\mu_1^2+\lambda_1 (Q_1+\lambda_1)-Q_2 \mu_1
+\frac{\lambda_1}{\lambda_2} \mu_1 (Q_1+2\lambda_1)+\mu_1^2
\frac{\lambda_1^2}{\lambda_2^2}.
\label{Feigen}
\eeqa
The accelerated expansion can be realized for 
\be
\xi<2\lambda_2^2\,,
\ee
in which case $\gamma_1, \gamma_2,\gamma_3$ are negative. 
The eigenvalue $\gamma_4$ is negative for 
\be
\left( \mu_1+\lambda^2/\lambda_2 \right)
\left( \mu_1-\mu_2 \right)<0\,.
\ee
This inequality is opposite to that of the stability condition (\ref{ca2}) 
derived for the fixed points A and B in the absence of the couplings $Q_i$.
We also require $\gamma_5<0$ to ensure the stability of the point F.

For the model parameters used in Fig.~\ref{fig3} all the eigenvalues (\ref{Feigen})
are negative, so the point F is in fact stable. 
Since $y_2=0$ at the point F, the potential 
$V_2 e^{-\lambda_1 (1+\mu_2/\lambda_2)\phi_1+\mu_2 \phi_2}$ 
does not contribute to the dynamics (see the evolution of $y_2^2$ in Fig.~\ref{fig3}).
The acceleration of the Universe is driven by the potential 
$V_1 e^{-\lambda_1 (1+\mu_1/\lambda_2)\phi_1+\mu_1 \phi_2}$. 
Numerically we confirmed that the asymptotic values of 
$w_x$, $w_{\rm eff}$, $x_1$, $x_2$, and $y_1^2$ show good 
agreement with the analytic estimation given above.
Depending on the model parameters $\lambda_1$, $\lambda_2$, $\mu_1$, 
$\mu_2$, $Q_1$, and $Q_2$, the attractor solutions are different.
Evaluating the eigenvalues (\ref{eigenA}) and (\ref{Feigen}) in each case, 
we readily know the stability of the fixed points A and F.

Finally we briefly discuss the case in which the point B may be responsible 
for the scaling attractor ($\Omega_x \simeq 0.7$) with cosmic acceleration.
As we mentioned above, if the scaling solution B is stable, 
the fixed point F is not stable. 
Setting $w_m=0$ in Eq.~(\ref{Omex}), the condition $\Omega_x<1$ 
translates to $q>3/\lambda^2-1$. Moreover the accelerated expansion 
is realized for $w_{\rm eff}=-q/(1+q)<-1/3$, that is, $q>1/2$.
These two conditions can be written as
\be
\frac{Q_1}{\lambda_1}+\frac{Q_2}{\lambda_2}>
3 \left( \frac{1}{\lambda_1^2}+\frac{1}{\lambda_2^2} \right)-1\,,
\qquad
\frac{Q_1}{\lambda_1}+\frac{Q_2}{\lambda_2}>\frac12\,.
\label{Bexis}
\ee
On the other hand, the $\phi$MDE can be compatible with the CMB 
observations for 
$w_{\rm eff}=\Omega_x=2(Q_1^2+Q_2^2)/3 \ll 1$ \cite{Amendola}, i.e., 
\be
Q_1^2+Q_2^2 \ll 1\,.
\label{phiMDEexis}
\ee
For the existence of the scaling accelerated point B we require the large couplings 
$Q_1$ and $Q_2$ satisfying the conditions (\ref{Bexis}), 
but the presence of the acceptable 
$\phi$MDE demands the small couplings.
In particular we need small values of $\lambda_1$ and $\lambda_2$ to 
make $Q_1/\lambda_1+Q_2/\lambda_2$ large, but in this case the 
rhs on the first inequality of Eq.~(\ref{Bexis}) tends to be even larger.
Hence, as in the single-field case \cite{Quartin}, it is generally difficult to realize 
the sequence of the $\phi$MDE followed by the scaling solution B
with $\Omega_x \simeq 0.7$ and $w_{\rm eff}<-1/3$.

\section{Conclusions} 
\label{consec}

In general k-essence model with multiple scalar fields $\phi_i$ ($i=1,\cdots,N$),
we derived the Lagrangian for the existence of cosmological scaling solutions 
in the presence of a barotropic perfect fluid coupled to $\phi_i$.
The resulting Lagrangian is simply given by Eq.~(\ref{scalinglag}), where 
$g$ is an arbitrary function in terms of $Y_i=X_i e^{\lambda_1 \phi_1}$ and 
$Z_i=\phi_{i+1}-\lambda_1 \phi_1/\lambda_{i+1}$. 
Along the scaling solution, the scalar fields evolve as Eq.~(\ref{phiso}) 
with $Y_i$ and $Z_i$ constant. 
For canonical multiple scalar fields the scaling solution 
behaves as an effective single field $\sigma$ with the trajectory 
given by Eq.~(\ref{dotsigma}).

For the multi-field Lagrangian (\ref{scalinglag}) we obtained the autonomous
equations (\ref{x1})-(\ref{z1}) by introducing the dimensionless variables
$x_i=\dot{\phi}_i/(\sqrt{6}H)$ and $y=e^{-\lambda_1 \phi_1/2}/(\sqrt{3}H)$. 
There are two important fixed points with $y \neq 0$ 
for arbitrary functions of $g$.
One of them is the scalar-field dominated point A ($\Omega_x=1$) 
satisfying the conditions (\ref{sca1})-(\ref{weffA}). 
We showed that the assisted inflationary mechanism is present 
for the fixed point A. Even if each field does not lead to cosmic acceleration, 
the multiple fields can cooperatively do so by reducing the effective slope 
$\lambda$ defined by Eq.~(\ref{lamdef}). 
Provided $\lambda^2<2$, the point A can be responsible for the 
accelerated expansion.
Another fixed point B is the scaling solution ($\Omega_x=$\,\,constant generally 
different from 1) 
satisfying the conditions (\ref{weffscaling})-(\ref{wx}).
In the presence of the couplings between the fields and the background 
fluid ($q \neq 0$), the effective equation of state $w_{\rm eff}$ 
for the point B is generally different from $w_m$.

If the Lagrangian (\ref{scalinglag}) is specified to some form, 
we can show the existence 
of kinetically driven fixed points satisfying $y=0$.
For the function (\ref{gchoice}), which involves the case of $N$ 
canonical scalar fields, there exist the $\phi$MDE point C characterized 
by $x_i=\sqrt{6}Q_i/[3c_i(w_m-1)]$ and the kinetic point D with $\Omega_x=1$. 
Unlike the fixed points A and B, the quantities $y$ and $Z_i$ are not 
necessarily constant for C and D. 
In the presence of the couplings $Q_i$, the standard matter-dominated 
epoch can be replaced by the $\phi$MDE with 
$w_{\rm eff}=\Omega_x=\sum_{i=1}^N 2Q_i^2/(3c_i)$.

In Sec.~\ref{stasec} we studied the stability of the fixed points A and B
in the two-field system described by the Lagrangian (\ref{lagtwo}).
For the point A it is possible to carry out the general analysis 
without specifying any functional form of $g$. 
The stability of this scalar-field dominated solution is ensured 
under the conditions (\ref{Acon1}), (\ref{Acon2}), and (\ref{Acon3}).
The stability analysis for the point B is too cumbersome to be written 
in a general way. In the model of two canonical fields with the function 
(\ref{twocanonical2}), we showed that the scaling solution B is stable 
under the conditions (\ref{Bsta1}) and (\ref{Bsta2}) with the 
additional requirement (\ref{hZre}).
These conditions can be nicely interpreted as a geometric approach 
based on the rotation in the field space. 
In the case of two canonical fields with $Q_i=0$, we found that the 
point A is stable (unstable) when the point B is unstable (stable).

In Sec.~\ref{example} we discussed the cosmological dynamics 
for the two-field model (\ref{conlag}) in the presence of radiation and 
non-relativistic matter. In doing so, it is convenient to employ the 
variables $y_1$ and $y_2$ defined in Eq.~(\ref{y2def}) instead 
of $y$ and $Z_1$. In this model there exist a few more fixed 
points than A, B, C, D relevant to the scaling radiation/matter 
eras and the epoch of cosmic acceleration. 
These fixed points are summarized in Table \ref{table2}. 

Depending on the model parameters and initial conditions of the model 
(\ref{conlag}), there are several qualitatively different cases:
(i) the scaling radiation era (B1) followed by the stable scaling 
matter era (B2),
(ii) the scaling radiation (E1) and matter (E2) eras followed by 
the point A with cosmic acceleration, and 
(iii) the $\phi$MDE point C followed by the accelerated expansion 
driven by another point F. 
The points E1, E2, and F, which satisfy $y_2=0$, arise 
when the potential energy of the last term of Eq.~(\ref{conlag}) 
is negligibly small relative to the third term.
The cases (ii) and (iii) can be distinguished from 
each other by the different evolution of the dark energy 
equation of state $w_x$.

It will be of interest to put observational bounds on the viable 
parameter space of the model (\ref{conlag}) and its extended model
by using the data of supernovae type Ia, CMB, baryonic acoustic oscillations, 
and the BBN. The BBN bound should not be so restrictive
in the multi-field context because the field density parameter $\Omega_x$ 
in the radiation era can be smaller 
than that in the single-field case.
The study of matter density perturbations is also important to place 
constraints on the couplings $Q_i$ in the presence of the $\phi$MDE.
We leave these topics for a future work.

\section*{ACKNOWLEDGEMENTS}

We thank Luca Amendola for useful discussions.  This work is supported
by the Grant-in-Aid for Scientific Research from JSPS (Nos.\,24540287
(TC), 24540286 (ST)), by the Grant-in-Aid for Scientific Research on
Innovative Areas (No.~21111006 (ST)) and in part by Nihon University
(TC). Some of this work was done at the Yukawa Institute, during the
workshop YITP-X-13-03.


\end{document}